\documentclass[fleqn,usenatbib]{mnras} 

\usepackage{newtxtext,newtxmath}

\usepackage[T1]{fontenc}
\usepackage{ae,aecompl}

\usepackage{graphicx}	
\usepackage{amsmath}	

\usepackage{grffile}    

\def\Mdot{\hbox{${\dot M}$}}
\def\vinfty{\hbox{${v_{\infty}}$} \,}

\def\km{{\rm\thinspace km}}
\def\s{{\rm\thinspace s}}
\def\yr{{\rm\thinspace yr}}

\def\erg{{\rm\thinspace erg}}
\def\kmps{\hbox{${\rm\km\s^{-1}\,}$}}
\def\ergps{\hbox{${\rm\erg\s^{-1}\,}$}}
\def\Rsol{\hbox{${\rm\thinspace R_{\odot}}$}}
\def\Msol{\hbox{${\rm\thinspace M_{\odot}}$}}
\def\Lsol{\hbox{${\rm\thinspace L_{\odot}}$}}
\def\Msolpyr{\hbox{${\rm\Msol\yr^{-1}\,}$}}
\def\spose#1{\hbox to 0pt{#1\hss}}
\def\ltsimm{\mathrel{\spose{\lower 3pt\hbox{$\sim$}}
        \raise 2.0pt\hbox{$<$}}}
\def\gtsimm{\mathrel{\spose{\lower 3pt\hbox{$\sim$}}
        \raise 2.0pt\hbox{$>$}}}

\title[Particle acceleration in colliding-wind binaries] 
{Particle acceleration and non-thermal emission in colliding-wind binary systems}

\author[J.~M.~Pittard et al.]{J.~M.~Pittard$^{1}$\thanks{E-mail:
    j.m.pittard@leeds.ac.uk}, G.~E.~Romero$^{2}$ and G.~S.~Vila$^{2}$\thanks{Currently at
    CONICET in YPF Tecnolog\'{i}a S.A.}\\
$^{1}$School of Physics and Astronomy, University of
       Leeds, Woodhouse Lane, Leeds LS2 9JT, UK\\
$^{2}$Instituto Argentino de Radioastronom\'{i}a, CCT-La Plata, CONICET, 1900FWA, La Plata, Argentina\\
}

\date{Accepted 2021 April 13. Received 2021 April 08; in original form
  2020 August 25}

\pubyear{2021}

\begin{document}
\label{firstpage}
\pagerange{\pageref{firstpage}--\pageref{lastpage}}
\maketitle

\begin{abstract}
  We present a model for the creation of non-thermal particles via
  diffusive shock acceleration in a colliding-wind binary. Our model
  accounts for the oblique nature of the global shocks bounding the
  wind-wind collision region and the finite velocity of the scattering
  centres to the gas. It also includes magnetic field amplification by
  the cosmic ray induced streaming instability and the dynamical back
  reaction of the amplified field. We assume that the injection of the
  ions and electrons is independent of the shock obliquity and that
  the scattering centres move relative to the fluid at the Alfv\'{e}n
  velocity (resulting in steeper non-thermal particle
  distributions). We find that the Mach number, Alfv\'{e}nic Mach
  number, and transverse field strength vary strongly along and
  between the shocks, resulting in significant and non-linear variations in the
  particle acceleration efficiency and shock nature (turbulent
  vs. non-turbulent). We find much reduced compression
  ratios at the oblique shocks in most of our models compared to our
  earlier work, though total gas compression ratios that
  exceed 20 can still be obtained in certain situations. We also
  investigate the dependence of the non-thermal emission on the
  stellar separation and determine when emission from secondary
  electrons becomes important. We finish by applying our model to
  WR\,146, one of the brightest colliding wind binaries in the radio
  band. We are able to match the observed radio emission and find that
  roughly 30 per cent of the wind power at the shocks is channelled
  into non-thermal particles.
\end{abstract}




\section{Introduction}
\label{sec:intro}
Colliding-wind binary (CWB) systems typically consist of two
early-type stars whose individual winds collide at supersonic speeds
\citep*[e.g.,][]{Stevens:1992,Pittard:2009}. This
interaction produces a wind-wind collision region (WCR) where strong
global shocks slow the winds and heat the plasma up to temperatures of
$10^{7}$\,K or more. The WCR may radiate strongly at X-ray energies,
the most famous examples perhaps being WR\,140
\citep[e.g.,][]{Pollock:2005,Sugawara:2015} and $\eta$\,Carinae
\citep[e.g.,][]{Hamaguchi:2007,Henley:2008,Corcoran:2010,Hamaguchi:2014a,Hamaguchi:2014b,Hamaguchi:2016,Hamaguchi:2018}. Numerical
simulations of the X-ray emission from the WCR have become
increasingly sophisticated in recent years
\citep[e.g.,][]{Pittard:2010,Parkin:2011a,Parkin:2011b,Parkin:2014}.

Particles may also be accelerated to high energies at the global
shocks bounding the WCR through diffusive shock acceleration
(DSA). The presence of such non-thermal particles is revealed by
synchrotron emission, which is sometimes spatially resolved
\citep*[e.g.,][]{Williams:1997,Dougherty:2000,Dougherty:2005,O'Connor:2005,Dougherty:2006,Ortiz-Leon:2011,Benaglia:2015,Brookes:2016}
or shows orbital variability
\citep[e.g.,][]{Blomme:2013,Blomme:2017}. Catalogues of
particle-accelerating CWB systems have been assembled by
\citet{DeBecker:2013} and \citet{DeBecker:2017}.

The convincing detection of
non-thermal emission at X-ray and $\gamma$-ray energies has proved far
more difficult, but at last this appears to be changing. In 2018,
non-thermal X-ray emission was reported from $\eta$\,Carinae
\citep{Hamaguchi:2018}. Crucially, the detection was made using
NuSTAR, a focusing telescope, which localised the emission to within a
few arc-seconds of the binary and revealed that the emission varied
with the orbital phase. This work provided much needed confirmation of
previous GeV detections \citep[e.g.,][]{Reitberger:2015} which
suffered from poor localisation. Most recently, $\eta$\,Carinae has
been detected at energies of 100's GeV by the HESS telescope
\citep{HESS:2020}.

Driven by these observations we are developing a numerical model for
simulating the non-thermal emission from CWBs. In this paper we take
the model presented in \citet*{Pittard:2020} and improve it in several
ways. Firstly, the particle acceleration scheme is updated to that in
\citet{Grimaldo:2019}, which generalizes \citet*{Caprioli:2009}'s
model for the case of oblique shocks where the background magnetic
field has also a transverse component. This scheme self-consistently
includes magnetic field amplification due to the cosmic ray induced
streaming instability, and the dynamical back reaction of the
amplified magnetic field. The back reaction reduces the modification
of the shock precursor and the total compression ratio of the shock,
compared to standard non-linear DSA. However, we improve
\citet{Grimaldo:2019}'s model by also considering the finite velocity
of the scattering centres relative to the fluid. This can have a big
effect on the steepness of the non-thermal particle distributions.

Secondly, we include a model for the magnetic field in the stellar
winds. Two possible configurations are considered: radial (applicable for non-rotating
stars) or toriodal (applicable for rotating stars). Thirdly, the creation of
secondary particles from proton-proton interactions is also
taken into account. Finally, additional emission and absorption processes are
modelled: synchrotron emission, two-photon absorption from the
creation of electron-positron pairs, and free-free absorption from the
clumpy winds.  In Sec.~\ref{sec:model} we describe our new model and
in Sec.~\ref{sec:results} we present the results. We apply our model
to the radio bright CWB WR\,146 in Sec.~\ref{sec:wr146} and we
summarize and conclude in Sec.~\ref{sec:summary}. Further details of
some of the improvements to the model are described in a set of Appendices.

\begin{table}
\begin{center}
\caption[]{The stellar parameters used in our standard model. Both stars are assumed to have an
effective temperature $T=40,000$\,K, a surface magnetic field flux
density $B_{*}=100$\,G, and an equatorial rotational speed of
$200\,\kmps$. Both winds are assumed to be clumpy with a volume
filling factor $f=0.1$. The stellar separation, $D=2\times10^{15}\,$cm.}
\label{tab:standard_parameters}
\begin{tabular}{lll}
\hline
Parameter & WR-star & O-star\\
\hline
$\Mdot\,\,(\Msolpyr)$ & $2\times10^{-5}$ & $2\times10^{-6}$ \\
$v_{\rm \infty}\,\,(\kmps)$ & $2000$ & $2000$ \\
$L\,\,(\Lsol)$ & $2\times10^{5}$ & $5\times10^{5}$\\
\hline
\end{tabular}
\end{center}
\end{table}

\section{The model}
\label{sec:model}

\subsection{Global structure and upstream quantities}
Our model is based on the one presented in \citet{Pittard:2020}. The
reader is referred to this paper for full details, but in brief it is
assumed that the stellar winds collide at fixed speeds to create an
axisymmetric WCR. Orbital effects and the acceleration/deceleration of
the winds are ignored, so our models are currently most appropriate
for wide binaries with long orbital periods where these neglected
effects are minimised. We also assume that the global shocks are
coincident with the contact discontinuity (CD) between the winds,
which is a suitable first-order approximation\footnote{\citet{Pittard:2018}
determined that the shocks flare away from the CD at angles of
$\approx20^{\circ}$ when the WCR is largely adiabatic.}. The position
of the CD is computed using the equations in \citet*{Canto:1996}, which
gives an accurate determination of the half-opening angle for wind
momentum ratios $>0.01$ \citep{Pittard:2018}.

Particle acceleration at the shock depends strongly on the assumed pre-shock magnetic
field. Close to each star the magnetic field is a dipole; it changes
to a radial configuration at distances beyond the Alfv\'{e}n radius,
$r_{\rm A}$, and, if the star is rotating, the field lines wrap up and
the field becomes toroidal at distances
$r > R_{*}v_{\infty}/v_{\rm rot}$, where $v_{\rm rot}$ is the
equatorial rotation speed of the star, $\vinfty$ is the wind speed and
$R_{*}$ is the stellar radius \citep[][]{Eichler:1993}. The
radial field is
\begin{equation}
B(r) = B_{*}\left(\frac{R_{*}}{r}\right)^{2},
\end{equation}
where $B_{*}$ is the magnetic flux density at the stellar surface. For the toroidal field we adopt
\begin{equation}
B_{\phi}(r,\theta) = B_{*}\frac{v_{\rm rot}}{v_{\infty}}\left(\frac{R_{*}}{r}\right)^{2}\left(\frac{r}{R_{*}}-1\right)\sin\theta,
\end{equation}
where $\theta$ is the polar angle, and $B_{\rm r} = B_{\theta} = 0$
\citep[see, e.g.,][]{Garcia-Segura:1997}. We adopt $B_{*}=100\,$G and
$v_{\rm rot}/v_{\infty} = 0.1$ as reasonable values \citep[see,
e.g.,][]{Eichler:1993}.  

Starting at the apex of the WCR the CD is divided into a sequence of
annuli of 1 degree interval in the angle $\theta$ measured from the
secondary star \citep[hereafter assumed to be the star with the less
powerful wind - see, e.g., Fig.~1 in][]{Pittard:2020}. Each annulus is
then subdivided into 8 segments equally spaced in azimuthal angle
$\Phi$, which measures the position on the WCR relative to the
rotation axis of each star (the latter are assumed to be aligned with
the orbital axis). $\Phi=0$ points upwards, while $\Phi=\pi/2$ lies in
the orbital plane. $\Phi$ increases in a clockwise direction for each
star. Therefore, particular values for $\theta$ and $\Phi$
correspond to a particular position on the WCR for each star (although
given the definition of $\Phi$ the points on each shock will be in
opposite halves of the model). The centre of each segment has
$\Phi = (2n-1)\pi/8$ where $n=1...8$\footnote{In some of the following figures
the shock properties and particle distributions are given at other
specific values of $\Phi$.}.

At the centre point of each segment the pre-shock wind properties are
calculated: the density, $\rho_{\rm 0}$, the velocity parallel
($u_{\rm 0\parallel}$) and perpendicular ($u_{\rm 0\perp}$) to the CD,
and the magnetic field flux density $B_{\rm 0}$ and angle to the shock
normal $\theta_{\rm B0}$. We set the pre-shock gas temperature to
$T_{\rm 0}=10^{4}\,$K, as appropriate for photoionized stellar winds.

\subsection{The shock solution}
The non-thermal particle spectrum at the shock is calculated by
solving the diffusion-advection equation, as detailed in
Appendix~\ref{sec:appDSA}, which provides all quantities of interest. The shock has a
precursor and a subshock. All of the far upstream quantities have a
subscript ``0''. Those immediately upstream of the subshock have a
subscript ``1'', while the postshock quantites have a subscript ``2''.
In solving the diffusion-advection equation we assume that all
quantities change locally only in the $x$-direction which is
perpendicular to the shock and that the magnetic field lies in the
$x$-$z$ plane.

Four compression factors are of interest. The first two relate to the
gas and are $R_{\rm tot} = u_{0x}/u_{2x}$ and
$R_{\rm sub} = u_{1x}/u_{2x}$. The second two relate to the scattering
centres and are $S_{\rm sub}$ and $S_{\rm tot}$. The non-thermal
particles produce turbulence created by resonant and non-resonant
instabilities. If the turbulence is assumed to be Alfv\'{e}n waves
(produced by resonant instabilities) the scattering center speed is
the Alfv\'{e}n speed. However, the nature of the turbulence created by
non-resonant cosmic ray current-driven instabilites is significantly
different to Alfv\'{e}n waves, and is not necessarily well described
in such terms. Using a Monte Carlo simulation of DSA,
\citet{Bykov:2014} found that the velocity of the scattering centres
relative to the fluid was significantly below the Alfv\'{e}n speed.
However, this is a complicated issue, that might well depend on the
level of turbulence upstream, something that is expected to be high in
line-driven stellar winds. Therefore, for the time being, we continue
to make the standard assumption that the scattering centres move
relative to the fluid at the Alfv\'{e}n velocity, $\nu_{\rm A}$. The
compression ratios experienced by the scattering centres are then
\begin{equation}
\label{eq:SsubStot}
S_{\rm sub} = \frac{u_{1x}-v_{\rm A1}}{u_{2x} + v_{\rm
    A2}}\hspace{0.5cm}{\rm and}\hspace{0.5cm}S_{\rm tot} = \frac{u_{0x}-v_{\rm A0}}{u_{2x} + v_{\rm
    A2}}.
\end{equation}

The non-thermal proton distribution function,
$f_{\rm p}(p) = f_{1}(p) = f_{2}(p)$, at each position on the shocks
is obtained (see Eq.~\ref{eq:f1p}). Various pressures are also
obtained: the gas thermal pressure, the gas ram pressure, the
cosmic-ray pressure, the pressure of the uniform magnetic field, and
the pressure of the magnetic waves. The solution also reveals the
fraction of the input energy flux that goes into cosmic rays that are
either advected downstream or escape upstream (see
Eq.~\ref{eq:energyFractions}). The sum of these fractions gives the
total fraction going into cosmic rays.

\subsubsection{Shock obliquity and particle injection}
\label{sec:obliquity}
The 1D kinetic treatment developed by Blasi and collaborators
\citep{Blasi:2002,Amato:2005,Amato:2006,Caprioli:2009} assumes that
the shock is parallel, with the upstream magnetic field aligned with
the shock normal. \citet{Grimaldo:2019} modified their solution to
include a pressure term for the uniform background field, but did not
consider how the DSA efficiency changes with the obliquity of the
shock. This is a fundamental issue that unfortunately is not yet fully
resolved. On the one hand, simulations by \citet{Caprioli:2014} using
a hybrid particle-in-cell code show that ion acceleration becomes very
inefficient for shock obliquities
$\theta_{\rm B0} \gtsimm 45^{\circ}$, as very few of the reflected
ions are able to move further upstream to be injected into the DSA
process \citep*{Caprioli:2015}. On the other hand,
\citet{Reville:2013} suggested that there exists a quasi-universal
shock behaviour, whatever the orientation of the far upstream field,
because the field in the immediate upstream region becomes completely
disordered. In addition, \citet*{vanMarle:2018} find efficient DSA for
large shock obliquities because the shock becomes corrugated (though
\citet{Haggerty:2019} disagree with these findings). Furthermore,
large ($\delta B/B_{0}$) Alfv\'{e}nic turbulence upstream may allow
injection of ions at shocks that are perpendicular on average
\citep{Giacalone:2005}, although for this process to be efficient the
fluctuations must be strong on length-scales comparable to the
gyroradius of suprathermal particles \citep*[see the discussion
in][]{Caprioli:2018}.

The injection of electrons into the DSA process has long been an
outstanding problem but this is beginning to be tackled through
simulations that now show the simultaneous acceleration of both
electrons and ions \citep*{Park:2015,Kato:2015}. In quasi-parallel
shocks, electrons are not efficiently accelerated and the fraction of
the shock energy that goes into them, $\zeta_{\rm e}
\ltsimm 10^{-3}$. On the other hand, \citet*{Xu:2020} find
that electrons are efficiently injected and accelerated via
shock-drift acceleration and then DSA in quasi-perpendicular shocks
($\theta_{\rm B0} = 63^{\circ}$) if $M$ and (particularly) $M_{\rm A}$
are high enough. In such cases $\zeta_{\rm e} \sim 0.1$. Thus
it might be the case that some shocks preferentially accelerate
electrons and not ions.

Faced with this current understanding there are two extreme positions
that can be taken with any model:
\begin{enumerate}
\item If the flow at the shock is strongly turbulent on small scales
  (perhaps because of turbulence far upstream), or if the
  shock becomes corrugated, then the electron and ion acceleration
  efficiency may be independent of the shock obliquity, with ions
  accelerated more efficiently than electrons.
\item If not, then quasi-perpendicular shocks may accelerate electrons
  efficiently but not ions (with no significant shock modification),
  while quasi-parallel shocks may accelerate ions efficiently (with
  significant shock modification) but not electrons.
\end{enumerate}
In the current work we adopt the former scenario in which the ion
acceleration efficiency is dependent on $M$, $M_{\rm A}$ and the
maximum momentum of the non-thermal protons, $p_{\rm max}$, but not on
$\theta_{\rm B0}$ (in particular, the value of $\chi$ in
Eq.~\ref{eq:eta} is kept fixed and independent of $\theta_{\rm
  B0}$). This might be consistent with the known clumpy nature of line-driven
stellar winds. In future work we will explore the second possibility.

\subsubsection{Maximum proton momentum}
\label{sec:pmax}
The solution to the diffusion-advection equation depends on $p_{\rm max}$, which
generally depends on geometrical \citep{Hillas:1984} or temporal
\citep{Lagage:1983} conditions\footnote{If the escaping cosmic rays are able to self-confine by
  creating upstream magnetic turbulence, the maximum cosmic ray energy
  may become independent of the strength of the ambient magnetic
  field, and instead depend on the time taken for the magnetic field
  to be amplified \citep{Bell:2013}. This possibility is not
  considered in our model.}. In exceptional circumstances $p_{\rm
  max}$ can be set by proton-proton losses \citep[as occurs in
$\eta$\,Carinae;][]{White:2020}, but this is not important in the
models in the current work where we find the geometrical condition
dominates. $p_{\rm max}$ is thus set by the
diffusion (escape) of particles from the shock, where the diffusion
length $l_{\rm diff} = r_{\rm shk}/4$, and where $r_{\rm shk}$ is the
distance of the shock from the star. This gives a maximum proton
energy $E_{\rm max}=l_{\rm diff} eB_{\rm 0} u_{\rm 0\perp}/c$, where
$e$ is the proton/electron charge. As in \citet*{Ellison:2004}, a
turnover is applied to the distribution at the highest energies,
according to
\begin{equation}
\label{eq:alphaCut}
\exp\left[-\frac{1}{\alpha}\left(\frac{p}{p_{\rm max}}\right)^{\alpha}\right],
\end{equation}
where $\alpha$ is a constant. Steeper turndowns are achieved with
higher values of $\alpha$, and in this work we adopt $\alpha=4$
\citep[$\alpha=1$ was used in][]{Pittard:2020}. This change was
necessary in order for the synchrotron emission to fall below the
thermal X-ray flux in our model of WR\,146 (see
Sec.~\ref{sec:wr146}). Values of $\alpha$ as high as 4 have been previously used in the
literature \citep[see][]{Ellison:2004}.

\subsubsection{Maximum electron momentum}
The non-thermal electron distribution function, $f_{\rm e}(p)$, is not
an output from the solution to the diffusion-advection equation.  In
keeping with usual practice we set
$f_{\rm e}(p) = K_{\rm ep}\,f_{\rm p}(p)$, with an exponential cut-off
at $p_{\rm max,e}$ that is set by radiative losses. We assume that
$K_{\rm ep} = 0.01$ \citep[as in][]{Pittard:2006b,Pittard:2020}.
Though the value of $K_{\rm ep}$ is not yet well measured or
constrained in CWBs, the value we use is consistent with the
well-established ratio of proton to electron energy densities for
Galactic cosmic rays of $\sim100$ \citep[][]{Longair:1994}, and is in
rough agreement with observations of young Galactic SNRs
\citep[e.g.,][]{Morlino:2012} and simulations
\citep[e.g.,][]{Park:2015} that find
$K_{\rm ep} \approx 10^{-3}-10^{-2}$.

The non-thermal electron distribution is also prevented from exceeding
the Maxwell-Boltzmann thermal distribution, which can occur when the
former has a very steep slope. In such cases we locally reduce the
value of $K_{\rm ep}$ and the normalization of $f_{\rm e}(p)$ so
that it matches and smoothly connects to the peak of the
Maxwell-Boltzmann thermal distribution \citep[cf. Fig.~1
in][]{Caprioli:2010}.

The value of $p_{\rm max,e}$ is calculated by balancing the local acceleration and
loss rates. For $\gamma \gg 1$, the electron acceleration rate is
given by \citep[see Eq.~6 in][]{Pittard:2006a}
\begin{equation}
\label{eq:dgammadtElectrons}
\frac{d\gamma_{\rm e}}{dt} = \frac{8}{9}\frac{R_{\rm sub}-1}{R_{\rm
      sub}}\frac{u^{2}}{c^{2}}\frac{eB}{m_{\rm e}c},
\end{equation}
where $u$ is the shock velocity ($u=u_{\rm 0\perp}$), and we have
assumed that the non-thermal electrons feel a compression factor
$R_{\rm sub}$\footnote{Electrons confined to the sub-shock will feel a
  compression ratio given by $S_{\rm sub}$, but those with higher
  energy will stream further upstream and downstream and will feel
  slightly greater compression. Our choice of $R_{\rm sub}$ is
  supposed to mimic this as often $R_{\rm sub}$ is slightly greater
  than $S_{\rm sub}$.}. We further assume that $u$ and $B$ take their
{\em pre-subshock} values ($u = u_{1}$ and $B=B_{1}$).  The loss rates are
given in Appendix~A of \citet{Pittard:2020}. For the synchrotron loss
rate we use the {\em total} (normal plus turbulent) {\em post-shock}
magnetic field i.e.
$B = B_{\rm 2,tot} = \sqrt{B_{2}^{2} + (\delta B_{2})^{2}}$, where
$B_{2}$ and $\delta B_{2}$ are the magnetic flux densities of the
postshock uniform and turbulent fields, respectively. This yields a
slightly lower maximum energy for the primary non-thermal electrons
and their emission, compared to using only the non-turbulent component
of the magnetic flux density.  The total magnetic flux density is also
used for the downstream synchrotron cooling and emission.

At relatively close stellar separations (e.g., $D_{\rm sep}
\ltsimm 10^{14}\,$cm), $p_{\rm max,e}$ will typically be set by inverse
Compton losses which will occur in the Thomson limit. However, in
wider systems the inverse Compton losses will be reduced by the
Klein-Nishina effect so that synchrotron cooling may become the
dominant energy loss mechanism. The cooling time as a function of
electron energy for various processes is shown in Fig.~6 of \citet{Pittard:2020}.

\subsection{The downstream solution}

The non-thermal particle spectrum downstream of the shock is
calculated by solving the kinetic equation. For a volume co-moving
with the underlying thermal gas, and ignoring diffusion and particle
escape, the energy distribution $n \equiv dN/dE$ as a function of time
$t$ and energy $E$ is given by the continuity equation
\citep*{Ginzburg:1964,Blumenthal:1970}:
\begin{equation}
\label{eq:kinetic}
\frac{\partial n(E,t)}{\partial t} + \frac{\partial
  (\dot{E}n(E,t))}{\partial E} = Q(E,t).
\end{equation}
The second term is an advection term in energy space due to cooling
processes and this equation is valid when the energy losses can be
treated as continuous \citep[for further details see Appendix~A
in][]{Pittard:2020}. $Q(E,t)$ is the number of particles per unit
volume injected in a time $dt$ in the energy range ($E,E+dE$); the
addition of this term marks an important difference from our previous
work \citep[cf.][]{Pittard:2020}, where we did not include it. Our
calculation of the injection function for secondary electrons
(actually electron-positron pairs), $Q_{\rm e \pm}$, is detailed in
Appendix~\ref{sec:appProtonProton}. The secondary electrons are
produced via the decay of charged pions, which are created in
collisions between thermal and non-thermal protons. The creation of
secondary electrons by interactions between photons and non-thermal
protons is detailed in Appendix~\ref{sec:appProtonPhoton}. We show in
Appendix~\ref{sec:appPPvsPgam} that proton-proton interactions
dominate the emissivity in our standard CWB model, justifying our
omission of interactions between non-thermal protons and stellar photons.

The fluid properties are assumed to be constant within each segment on
the WCR. When the co-moving volume containing the thermal and
non-thermal particles moves along the CD and into the next segment the
fluid properties are set to those of the new segment. This leads to a
steady decrease in the gas density and temperature and the magnetic
flux density, and a steady increase in the velocity. The photon energy
density from the stars also drops. The number density of the
non-thermal particles is also reduced through the relative adiabatic
expansion between the two segments \citep[cf. Appendix~A
in][]{Pittard:2020}.

\subsection{Further details}
Our new model also includes synchrotron emission, and photon-photon
and free-free absorption. Details of the calculation of the
synchrotron emission are provided in
Appendix~\ref{sec:appSynchrotronEmission}. The absorption of high
energy photons by collisions with stellar photons to create
electron-positron pairs is detailed in
Appendix~\ref{sec:appPhotonPhotonAbsorption}\footnote{The emission
  from such pairs is currently not calculated. Nor do we consider the
  possibility of an inverse Compton pair cascade
  \citep*[e.g.,][]{Bednarek:2005,Khangulyan:2008}.  For an inverse
  Compton cascade to develop, inverse Compton energy losses must
  dominate over synchrotron energy losses \citep[as seen in Fig.~6
  of][]{Pittard:2020}. Assuming a toroidal magnetic field in the
  stellar winds, the spectrum up to TeV energies will be affected if
  the surface magnetic field,
  $B_{*} \ltsimm 5 \frac{L_{5}^{0.5}}{R_{10}}\frac{v_{\infty}}{v_{\rm
      rot}}$,
  where $L_{5}$ is the stellar luminosity in units of $10^{5}\,\Lsol$
  and $R_{10}$ is the stellar radius in units of $10\,\Rsol$. Note
  that there is no dependence on the distance to the star(s). With our
  standard model parameters we obtain $B_{*} \ltsimm 150$\,G. Thus
  pair cascades can be expected to develop in CWB systems when the
  optical depth for $\gamma$-$\gamma$ absorption becomes significant
  (which requires $D\ltsimm10^{13}\,$cm for our standard model
  parameters - see Fig.~\ref{fig:stanCWB_ntEmission_varDsep}).}. The
free-free absorption of low energy photons is detailed in
Appendix~\ref{sec:appFreeFreeAbsorption}. A final change to our
previous model is the use of the \citet*{Khangulyan:2014}
approximation for the inverse Compton emissivity when the target
photons have a black-body distribution \citep[see
also][]{delPalacio:2020}. This removes one loop from the calculations
and leads to a significant speed-up with no loss of accuracy.

\begin{figure*}
\includegraphics[width=17.5cm]{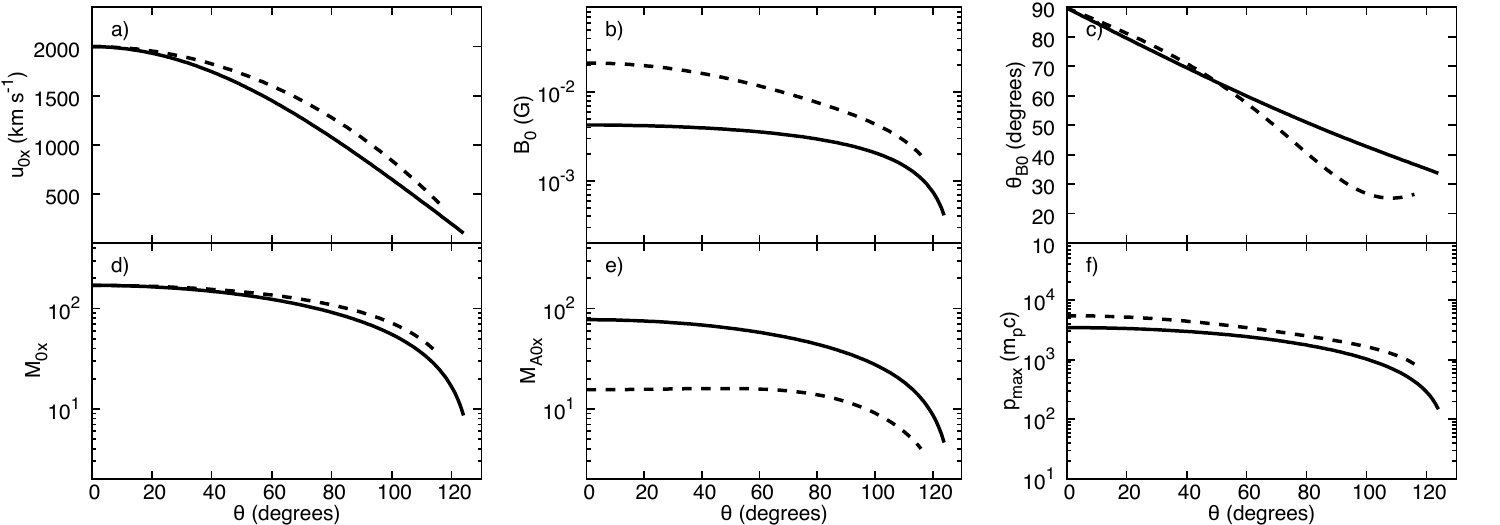}
\caption{Selected quantities along the CD as a function of the angle
  $\theta$ from the secondary star (see Fig.~1 in \citet{Pittard:2020}
  for the definition of $\theta$). Panels a-f) show the upstream
  perpendicular wind velocity ($u_{\rm 0x}$), magnetic flux density
  ($B_{0}$), angle of the magnetic field to the shock normal
  ($\theta_{\rm B0}$), perpendicular Mach number ($M_{\rm 0x}$),
  perpendicular Alfv\'{e}nic Mach number ($M_{\rm A0x}$), and the
  maximum non-thermal proton momentum ($p_{\rm max}$). In all panels
  the solid line indicates the properties for the WR-shock, while the
  dashed line indicates the properties for the O-shock. $B_{0}$,
  $\theta_{\rm B0}$, $M_{\rm A0x}$ and $p_{\rm max}$ are all dependent on the
  azimuthal angle of the position on the WCR. The values in this
  figure are for $\Phi = \pi/4$.}
\label{fig:stanCWB_shockstats1}
\end{figure*}

\begin{figure*}
\includegraphics[width=17.5cm]{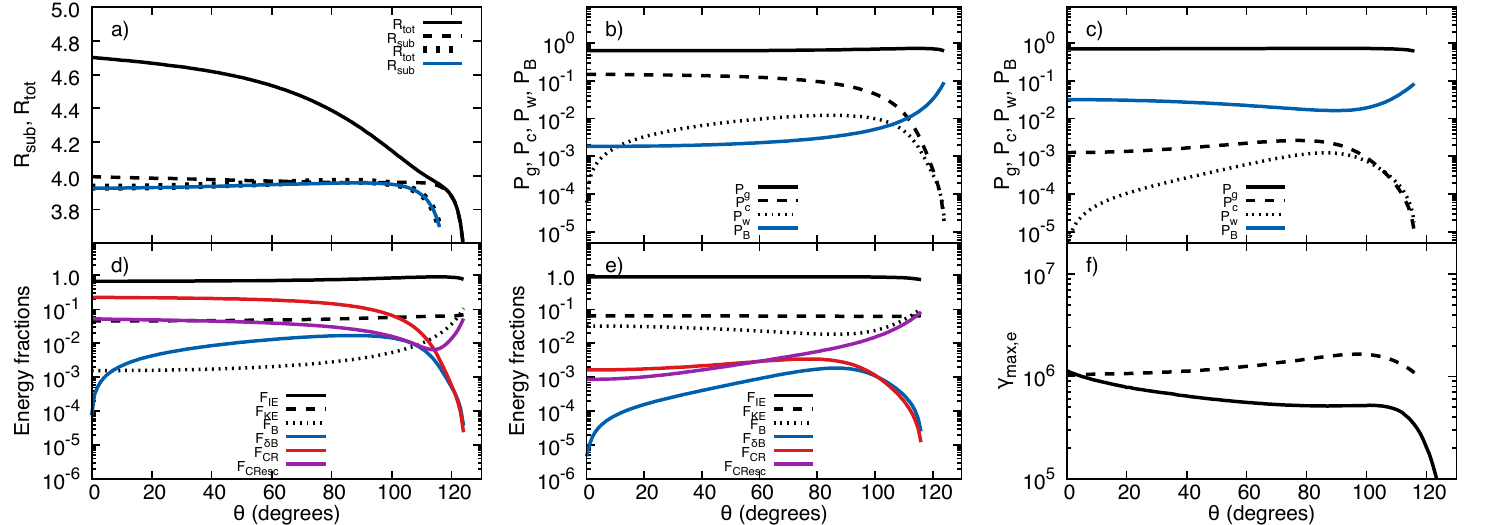}
\caption{Selected quantities along the CD as a function of the angle
  $\theta$ from the secondary star. Panel a) shows the gas compression
  ratio across the subshock ($R_{\rm sub}$) and the entire shock
  ($R_{\rm tot}$). The properties for the WR-shock are shown by the
  black solid and dashed lines. The properties for the O-shock are
  shown by the black dotted and blue solid lines. Panels b) and c)
  show the postshock pressure (normalized to $\rho_{0}u_{0x}^{2}$) for
  the gas ($P_{\rm g}$), cosmic rays ($P_{\rm c}$), magnetic
  turbulence ($P_{\rm w}$) and uniform magnetic field ($P_{\rm B}$).
  Panel b) shows the properties for the WR-shock while panel c) shows
  the properties for the O-shock. Panels d) and e) show the downstream
  thermal ($F_{\rm IE}$), kinetic ($F_{\rm KE}$), magnetic
  ($F_{\rm B}$), magnetic turbulence ($F_{\delta \rm{B}}$) and cosmic
  ray ($F_{\rm CR}$) energy fractions. $F_{\rm CResc}$ is the fraction
  of energy in the cosmic rays that escape upstream of the shock. The
  total energy fraction of the cosmic rays is
  $F_{\rm CR}+F_{\rm CResc}$. Panel d) shows the properties for the
  WR-shock while panel e) shows the properties for the O-shock. Panel
  f) shows the maximum Lorentz factor of the non-thermal electrons
  from the WR-shock (solid) and the O-shock (dashed) (the maximum
  proton momentum is shown in Fig.\ref{fig:stanCWB_shockstats1}). The
  values in this figure are for $\Phi = \pi/4$.}
\label{fig:stanCWB_shockstats2}
\end{figure*}

\begin{figure}
\includegraphics[width=8.0cm]{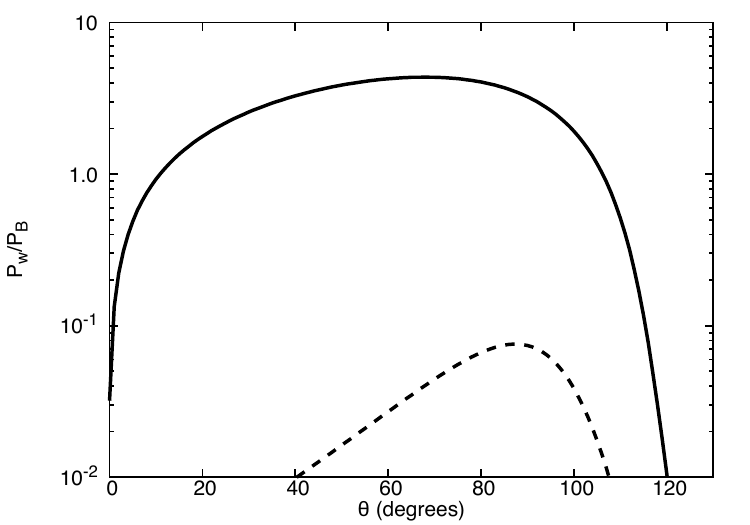}
\caption{Ratio of the turbulent to non-turbulent magnetic field as a
  function of angle (position) along the contact discontinuity. The
  properties of the WR-shock are shown by the solid line while the
  dotted line shows the properties of the O-shock. The post-shock
  magnetic field is turbulent for most of the WR-side of the CD, but
  is more ordered on the O-side.}
\label{fig:stanCWB_magneticTurbulence}
\end{figure}

\section{Results}
\label{sec:results} 
Unless otherwise noted we adopt a set of ``standard'' parameters for
our model, in which the stellar separation is $D = 2\times10^{15}$\,cm
and the viewing angle $\phi=90^{\circ}$ \citep[i.e. the line-of-sight
is perpendicular to the line-of-centres between the stars - see also
Fig.~1 in][]{Pittard:2020}. Other parameters of our model are noted in
Table~\ref{tab:standard_parameters}. The model is not of any
particular system, but its parameter values are chosen to be
representative of a WR+O system with a reasonably wide stellar
separation. For simplicity the DSA model assumes that the winds are
pure hydrogen, but all other parts of the code use WC mass fractions
($X=0.0$, $Y=0.5$, $Z=0.5$) for the WR-star and solar mass fractions
\citep[$X=0.7381$, $Y=0.2485$, $Z=0.0134$;][]{Grevesse:2010} for the
O-star. The wind momentum ratio is 0.1 and the stagnation point is at
a distance of $0.26\,D$ from the O-star, where $D$ is the stellar
separation. The WCR is largely adiabatic. Numerical values of
some pre-shock quantities are given in Sec.~2.7 of
\citet{Pittard:2020}.  With a toriodal magnetic field in the wind of
each star, the pre-shock magnetic flux density on the line of centres
is 4\,mG for the WR-shock and 20\,mG for the O-shock. The shocks are
almost perpendicular at this location.

\subsection{The standard model}
\subsubsection{Quantities along each shock}
\label{sec:stanCWB_shkstats}
Fig.~\ref{fig:stanCWB_shockstats1} shows various quantites from our
standard model as a function of angle, $\theta$, along the CD as
measured from the secondary star ($\theta=0^{\circ}$ corresponds to
the stagnation point of the WCR on the line-of-centres between the
stars, while $\theta=90^{\circ}$ indicates a point on the CD where
$z=D$ - see Fig.~1 in \citet{Pittard:2020}). The maximum value of
$\theta$ is $180^{\circ}$ minus the half-opening angle of the WCR. For
our standard parameters, the half-opening angle is
$\approx 50^{\circ}$, so $\theta_{\rm max} \approx 130^{\circ}$. 

Fig.~\ref{fig:stanCWB_shockstats1}a) shows that the perpendicular
pre-shock velocity is equal to $2000\,\kmps$ on the axis of symmetry
of the WCR (i.e. at the stagnation point between the winds), but
steadily declines as one moves off-axis. The WR-shock becomes
more oblique more rapidly than the O-shock. $u_{0x} = u_{\rm 0\perp} \rightarrow 0$ as
$\theta \rightarrow \theta_{\rm max}$.

Fig.~\ref{fig:stanCWB_shockstats1}b) shows that the pre-shock magnetic
flux density is significantly higher for the O-shock than for the
WR-shock. This mainly reflects the fact that the WCR is much closer to the
O-star, though there is also some enhancement due to the larger radius
of the O-star. This is a key difference to our previous work
\citep{Dougherty:2003,Pittard:2006a,Pittard:2006b,Pittard:2020} where
the on-axis pre-shock magnetic flux density was assumed to be
identical for both winds.

The angle that the pre-shock magnetic field makes to the shock normal,
$\theta_{\rm B0}$, is nearly $90^{\circ}$ near the stagnation point
(i.e. the shock is very nearly perpendicular - see
Fig.~\ref{fig:stanCWB_shockstats1}c). As one moves off-axis the field
becomes more oblique. Note that the value of $\theta_{\rm B0}$ is a
function of both $\theta$ and $\Phi$. Thus the particle acceleration
is no longer azimuthally symmetric if the stars are rotating and the
pre-shock magnetic field is toriodal. In such a case the shock is
always perpendicular when $\Phi=0$ or $\pi$ (for all values of
$\theta$). However, while it starts off perpendicular when
$\Phi=\pi/2$ (when $\theta=0^{\circ}$), it becomes more and more
oblique as $\theta$ increases.

Fig.~\ref{fig:stanCWB_shockstats1}d) shows that the pre-shock
perpendicular Mach number is above 100 for both shocks up to
$\theta \approx 90^{\circ}$. The pre-shock perpendicular Alfv\'{e}nic
Mach number is a factor of two lower than $M_{0x}$ for the WR-shock,
but for the O-shock $M_{A0x} \sim 0.1 M_{0x}$, so that
$M_{A0x} \ltsimm 15$ (Fig.~\ref{fig:stanCWB_shockstats1}e). As we will
see, this drastically affects the particle acceleration efficiency of
the O-shock in the standard model.

Fig.~\ref{fig:stanCWB_shockstats1}f) shows that the maximum
non-thermal proton momentum is nearly twice as high for the O-shock.
This is because the stronger magnetic field more than compensates for
the reduced radius of the shock.

While the focus of Fig.~\ref{fig:stanCWB_shockstats1} is mostly on
pre-shock quantities along each shock, the focus of
Fig.~\ref{fig:stanCWB_shockstats2} is mostly on the post-shock
quantities. It is immediately clear from the values of $R_{\rm tot}>4$
that the WR-shock is an efficient accelerator of non-thermal particles
(until $\theta \gtsimm 110^{\circ}$), while the O-shock is not (see
Fig.~\ref{fig:stanCWB_shockstats2}a). The reason for this is due to
the different values of $M_{\rm A0x}$ for the two shocks. When
$M_{\rm A0x}$ is small enough, the compression ratios felt by the
non-thermal particles ($S_{\rm sub}$ and $S_{\rm tot}$; see
Eq.~\ref{eq:SsubStot}) can become significantly lower than that felt
by the fluid ($R_{\rm sub}$ and $R_{\rm tot}$). This leads to a
steeper spectral slope for the non-thermal particles, as noted by
\citet{Bell:1978}. This is discussed further in Sec.~\ref{sec:particleDistributions}.

Fig.~\ref{fig:stanCWB_shockstats2}b) shows the various post-shock
pressures, normalized to the pre-shock ram pressure, for the
WR-shock. The thermal gas pressure still dominates but the cosmic ray
pressure remains above 10\% until $\theta > 80^{\circ}$.  The
accelerated particles are also able to generate significant magnetic
turbulence. Along a significant part of the WR-shock the turbulent
field exceeds the uniform field, by up to a factor of 2. In contrast,
the post-shock pressure from the non-turbulent magnetic field is the
second most important pressure behind the O-shock
(Fig.~\ref{fig:stanCWB_shockstats2}c), and the pressure from the
turbulent field is lower. This means that the post-shock magnetic
field is turbulent for most of the WR-side of the CD, but is more
ordered on the O-side (see Fig.~\ref{fig:stanCWB_magneticTurbulence}).

The WR-shock manages to convert 20\% of the input energy flux into
cosmic rays that are advected downstream
(Fig.~\ref{fig:stanCWB_shockstats2}d). A further 5\% goes into cosmic
rays that escape upstream. In contrast, the O-shock puts $< 1$\% of
the incoming energy flux into downstream cosmic rays (nearly 10\% goes
into cosmic rays that escape upstream as
$\theta \rightarrow \theta_{\rm max}$, but there is very little energy
going into the shock at this stage). Thus the non-thermal particles
accelerated at the WR-shock will dominate the non-thermal emission, as
we show in Sec.~\ref{sec:stanCWB_ntemission}.

Fig.~\ref{fig:stanCWB_shockstats2}f) shows that radiative losses limit
the maximum Lorentz factor of the non-thermal electrons to
$\gamma_{\rm max,e}\sim 10^{6}$ (for the protons $\gamma_{\rm max}
\gtsimm 10^{3}$). This is slightly lower than in
\citet{Pittard:2020}. Part of it is due to the stronger synchrotron
losses that are now assumed (i.e. the use of $B_{\rm 2,tot}$ rather
than $B_{0}$).  The change in $\gamma_{\rm max,e}$ with $\theta$
reflects the changing pre-shock magnetic field, compression, and
generation of the turbulent magnetic field.

\subsubsection{The shock precursor}
\begin{figure*}
\includegraphics[width=17.5cm]{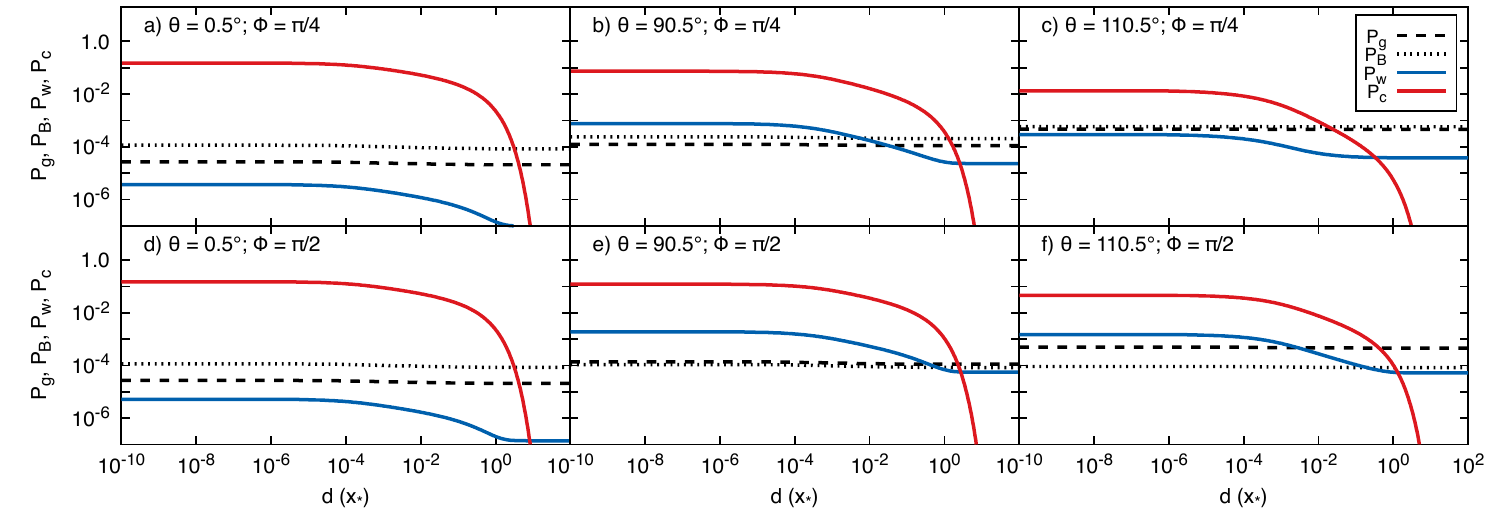}
\caption{The normalized thermal ($P_{\rm g}$), magnetic ($P_{\rm B}$), magnetic
  turbulence ($P_{\rm w}$), and cosmic-ray ($P_{\rm c}$)
  pressures in the WR-shock precursor. The subshock is located at the
  left side of each panel and the upstream flow is incident from the
  right. The location upstream of the subshock is in units of
  $x_{*}=D(p_{\rm max})/u_{0x}$. Each panel shows a different location
  on the WR-shock front, indicated by the values of $\theta$ and
  $\Phi$. In all cases the normalized perpendicular pre-shock velocity
  $U_{\rm x}$ remains close to 1.0 (not shown).}
\label{fig:stanCWB_precursor_shk0}
\end{figure*}

Fig.~\ref{fig:stanCWB_precursor_shk0} shows pressure
profiles in the WR-shock precursor as a function of position on the
WCR. Panels a)-c) are for $\Phi=\pi/4$ while panels d)-f) are for
$\Phi=\pi/2$ (i.e. positions on the WCR that lie in the orbital
plane). Panels a)-c) show that the normalized cosmic ray pressure drops
as $\theta$ increases, reflecting the drop in the acceleration
efficiency. Note that there is practically no difference between
panels a) and d), as the pre-shock magnetic flux density and angle to
the shock normal is almost identical. However, there are significant
differences between panels c) and f) since there are now much larger
differences in $B_{0}$ and $\theta_{\rm B0}$ at these positions on the
WR-shock. 

$U_{0x}$ is almost equal to 1.0 for all of the cases in
Fig.~\ref{fig:stanCWB_precursor_shk0} (the minimum value of
$U_{0x}=0.85$ is obtained for $\theta=0.5^{\circ}$). Finally,
Fig.~\ref{fig:stanCWB_precursor_shk0} shows that while the cosmic ray
and turbulent magnetic pressures both increase in the precursor, there
is very little increase in the thermal and magnetic pressures (again
this is most obvious when $\theta=0.5^{\circ}$), indicating that there
is little compression and/or heating within the precursor.

\begin{figure*}
\includegraphics[width=17.5cm]{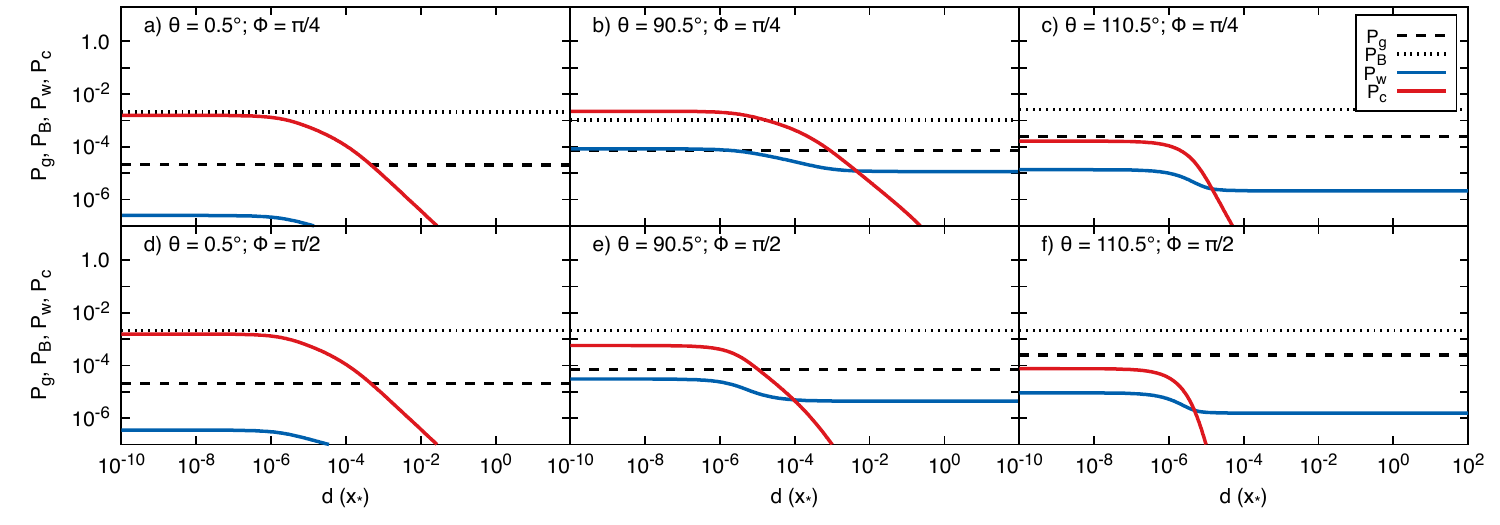}
\caption{As Fig.~\ref{fig:stanCWB_precursor_shk0} but for the
  O-shock. The precursor is less extended compared to that of the
  WR-shock (see Fig.~\ref{fig:stanCWB_precursor_shk0}).}
\label{fig:stanCWB_precursor_shk1}
\end{figure*}

Fig.~\ref{fig:stanCWB_precursor_shk1} shows the profiles in the
precursor of the O-shock. We see again that this shock is far less
effective at accelerating particles than the WR-shock, and also that
the precursor is less extended. We see that when $\Phi=\pi/4$, the
normalized cosmic ray pressure immediately upstream of the subshock,
$P_{\rm c1}$, is roughly constant for $\theta=0-90^{\circ}$, but drops
sharply for $\theta>90^{\circ}$ (panels a-c). However, when
$\Phi=\pi/2$, $P_{\rm c1}$ drops continuously with $\theta$.

Table~\ref{tab:xstar} notes the values of $x_{*}$ for each position on
the WCR shown in Figs.~\ref{fig:stanCWB_precursor_shk0}
and~\ref{fig:stanCWB_precursor_shk1}. Fig.~\ref{fig:stanCWB_precursor_shk0}
shows that the cosmic rays stream up to distances $\sim 10\,x_{*}$
from the WR-subshock, but are confined to distances $\ll x_{*}$ from
the O-subshock. We see that the cosmic ray precursor is generally much
smaller than the local scale of the shock \citep[taken to be
$r_{\rm WR}D$ and $r_{\rm OB}D$ for the on-axis ($\theta=0^{\circ}$)
WR and O-shocks, respectively - for the standard model
$r_{\rm WR}=0.74$ and $r_{\rm OB}=0.26$ - see also Fig.~1
in][]{Pittard:2020}. While the size of the WR-shock precursor relative
to the WCR starts to become significant as one moves off-axis, for the
most part our use of a 1-dimensional cosmic-ray shock model is valid
and appropriate. The far-off-axis region of the WCR adds little to the
total cosmic-ray population and emission, in any case.

\begin{table}
\begin{center}
\caption[]{The value of $x_{*}$ in our standard model, as a function
  of $\theta$ and $\Phi$. The stellar separation, $D = 2\times10^{15}$\,cm.}
\label{tab:xstar}
\begin{tabular}{llll}
\hline
$\theta$ ($^\circ$) & $\Phi$ ($^{\rm c}$) & \multicolumn{2}{c}{$x_{*}$ (cm)} \\
\hline
 & & WR-shock & O-shock \\
0.5 & $\pi/4$ & $2.7\times10^{13}$ & $1.0\times10^{13}$\\
0.5 & $\pi/2$ & $2.7\times10^{13}$ & $1.0\times10^{13}$ \\
90.5 & $\pi/4$ & $1.1\times10^{14}$ & $8.0\times10^{13}$ \\
90.5 & $\pi/2$ & $1.6\times10^{14}$ & $8.0\times10^{13}$ \\
110.5 & $\pi/4$ & $2.2\times10^{14}$ & $1.9\times10^{14}$ \\
110.5 & $\pi/2$ & $3.1\times10^{14}$ & $2.1\times10^{14}$\\
\hline
\end{tabular}
\end{center}
\end{table}

\subsubsection{The particle distributions}
\label{sec:particleDistributions}
\begin{figure}
\includegraphics[width=8.0cm]{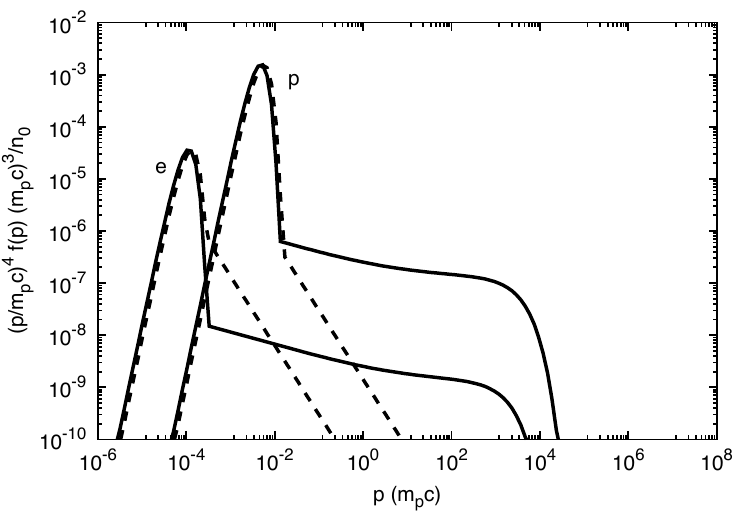}
\caption{The proton and electron distributions for the WR-shock (solid
  line) and O-shock (dashed line) for $\theta=0^{\circ}$ and
  $\Phi=\pi/4$. For the WR-shock, $R_{\rm tot} = 4.7$,
  $R_{\rm sub} = 3.99$, $S_{\rm tot} = 4.1$ and $S_{\rm sub} = 3.5$.
  For the O-shock, $R_{\rm tot} = R_{\rm sub} = 3.93$ and
  $S_{\rm tot} = S_{\rm sub} = 2.45$. For both shocks
  $n_{0} = 1.3\times10^{5}\,{\rm cm^{-3}}$. The thermal peaks are
  visible at low momenta.}
\label{fig:stanCWB_mtmSpectra}
\end{figure}

\begin{figure}
\includegraphics[width=8.0cm]{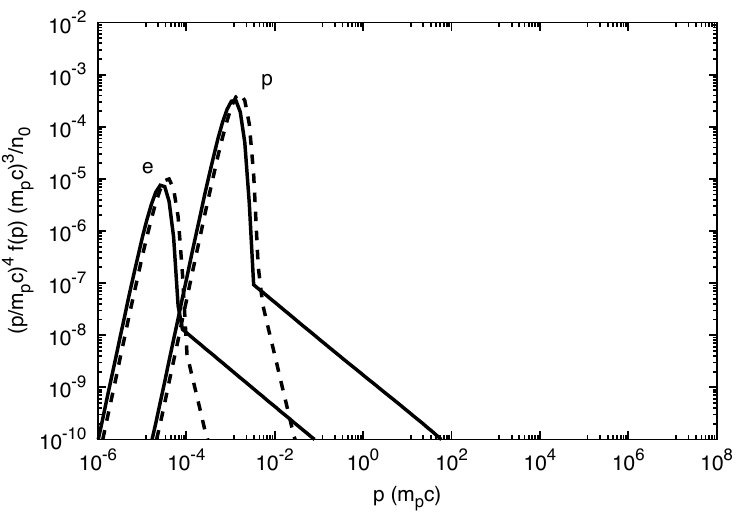}
\caption{The proton and electron distributions for the WR-shock (solid
  line) and O-shock (dashed line) for $\theta=110^{\circ}$ and
  $\Phi=\pi/4$. For the WR-shock, $R_{\rm tot} = 4.02$,
  $R_{\rm sub} = 3.96$, $S_{\rm tot} = 2.96$ and $S_{\rm sub} = 2.91$.
  For the O-shock, $R_{\rm tot} = R_{\rm sub} = 3.89$ and
  $S_{\rm tot} = S_{\rm sub} = 1.99$. For the WR-shock
  $n_{0} = 2.0\times10^{4}\,{\rm cm^{-3}}$, while for the O-shock
  $n_{0} = 4.2\times10^{3}\,{\rm cm^{-3}}$.}
\label{fig:stanCWB_mtmSpectra2}
\end{figure}

Figs.~\ref{fig:stanCWB_mtmSpectra} and~\ref{fig:stanCWB_mtmSpectra2}
show the distributions of the thermal and non-thermal particles
immediately downstream of the subshock. In each figure the proton
distributions are indicated by a ``p'', while the electron
distributions are indicated by an ``e''. The particle distributions
are shown for the WR-shock (solid line) and the O-shock (dashed
line). Fig.~\ref{fig:stanCWB_mtmSpectra} shows the distributions for
$\theta=0^{\circ}$, while Fig.~\ref{fig:stanCWB_mtmSpectra2} shows
them for $\theta=110^{\circ}$. In both cases $\Phi=\pi/4$.

The stand-out feature in both figures is the slope of the non-thermal
particle distribution. The spectral index $n$ of the particle
distribution ($f(p) \propto p^{n}$) is given by
\begin{equation}
\label{eq:spectralIndex}
n = -\frac{3r}{r-1},
\end{equation}
where $r$ is the relevant compression ratio. For a strong,
non-relativistic shock with polytropic index $\gamma=5/3$, the density compression
ratio is $r = R_{\rm tot} = 4$, which gives $n=-4$ (i.e. a flat
distribution in our figures). However, if the scattering centres move
relative to the fluid their compression ratio can be reduced, leading
to steeper spectra.  The on-axis WR-shock has $R_{\rm tot} = 4.7$,
$R_{\rm sub} = 3.99$, $S_{\rm tot} = 4.1$ and $S_{\rm sub} = 3.5$. The
spectral index of the particle distribution should therefore vary from
$n = -4.2$ at low energies to $n=-3.97$ at high energies, which is
indeed consistent with Fig.~\ref{fig:stanCWB_mtmSpectra} (the high
energy slope is not seen due to the maximum energy cut-off of the
particles).  For the on-axis O-shock we find
$R_{\rm tot} = R_{\rm sub} = 3.93$ but
$S_{\rm tot} = S_{\rm sub} = 2.45$. This yields $n=-5.1$ and is again
consistent with the displayed distribution.

The stellar parameters of our standard model are not too dissimilar from
those used by \citet{delPalacio:2016} to model HD\,93129A.  It is therefore
interesting that in order to match the observed synchrotron emission
from HD\,93129A, \citet{delPalacio:2016} adopt an energy index of
$p=3.2$ for the non-thermal particles in their model. This corresponds
to $n=-5.2$ for the momentum index of the particles, and is very
similar to the index that we find for the particles accelerated at the
on-axis O-shock (see Fig.~\ref{fig:stanCWB_mtmSpectra}).
\citet{delPalacio:2020} also consider a hardening of the high energy
particle distribution to $p=2$ (equivalent to $n=-4$), which is what
we obtain for the WR-shock in our model. 

In Fig.~\ref{fig:stanCWB_mtmSpectra} the curvature of the non-thermal
part of the distributions from the WR-shock reveal some modest shock
modification, but it is much reduced compared to the pure hydrodynamic
case \citep[cf. Fig.~4 in][]{Pittard:2020}, and this is also manifest
in the shift to higher momenta of the thermal
peaks. Fig.~\ref{fig:stanCWB_mtmSpectra} again indicates that the
particle acceleration at the O-shock is very inefficient.
Fig.~\ref{fig:stanCWB_mtmSpectra2} shows that as $\theta$ increases
the particle acceleration also becomes inefficient for the WR-shock,
with the observed steepening of the particle distribution consistent
with $S_{\rm sub}=2.91$ (giving $n\approx -4.6$). This behaviour again
contrasts with \citet{Pittard:2020} - see their Fig.~5.

\subsubsection{The non-thermal emission}
\label{sec:stanCWB_ntemission}
\begin{figure}
\includegraphics[width=8.0cm]{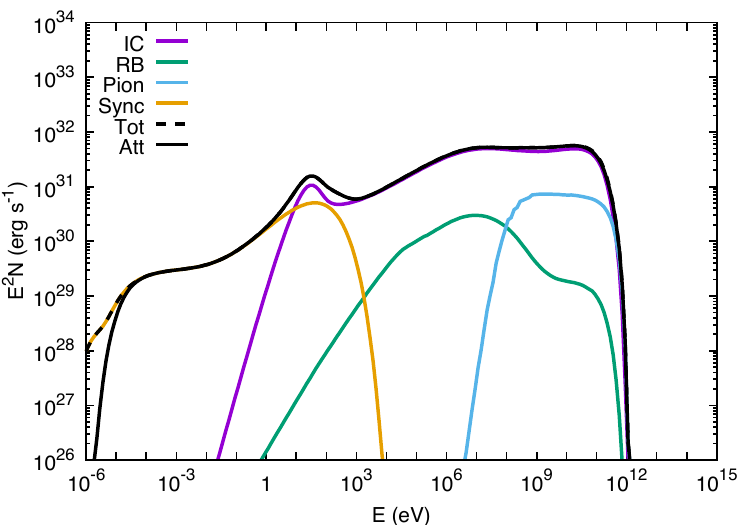}
\caption{The non-thermal emission from our standard model. The
  intrinsic emission from the inverse Compton (IC), relativistic
  bremmstrahlung (RB), $\pi^{0}$-decay (Pion), and the synchrotron
  (Sync) processes are shown separately, as well as their combined
  total (Tot) and the total attenuated (Att) emission (which accounts
  for free-free and photon-photon absorption). The stellar separation
  $D=2\times10^{15}\,{\rm cm}$ and the viewing angle
  $\phi=90^{\circ}$.}
\label{fig:stanCWB_ntEmission}
\end{figure}

The non-thermal emission from our standard model is shown in
Fig.~\ref{fig:stanCWB_ntEmission}. The inverse Compton emission
dominates at $E\gtsimm1\,$keV while synchrotron emission dominates for
$E \ltsimm 10$\,eV. Free-free absorption by the stellar winds causes
the synchrotron emission to turnover at about 2\,GHz \citep[which is
comparable to the turnover frequency found from the full hydrodynamic
models in][]{Pittard:2006a}. However, the Razin turnover frequency
occurs at about 5\,GHz and dominates the low frequency turnover in
this model. The emission from $\pi^{0}$-decay adds slightly more than
10\% to the total emission between $0.3-100$\,GeV. The non-thermal
emission in Fig.~\ref{fig:stanCWB_ntEmission} is somewhat softer than
that seen in Fig.~10 in \citet{Pittard:2020}, where a more pronounced
upwards curvature in the emission towards higher energies is
seen. This is due to the lower compression ratio of the scattering
centres which steepens the particle distributions in the current work,
and the less strongly modified shocks. The $\pi^{0}$-decay emission is
also weaker relative to the inverse Compton emission in the current
work, due to the lower total gas compression ratio $R_{\rm
  tot}$. $\gamma$-$\gamma$ absorption is negligible.

\begin{figure}
\includegraphics[width=8.0cm]{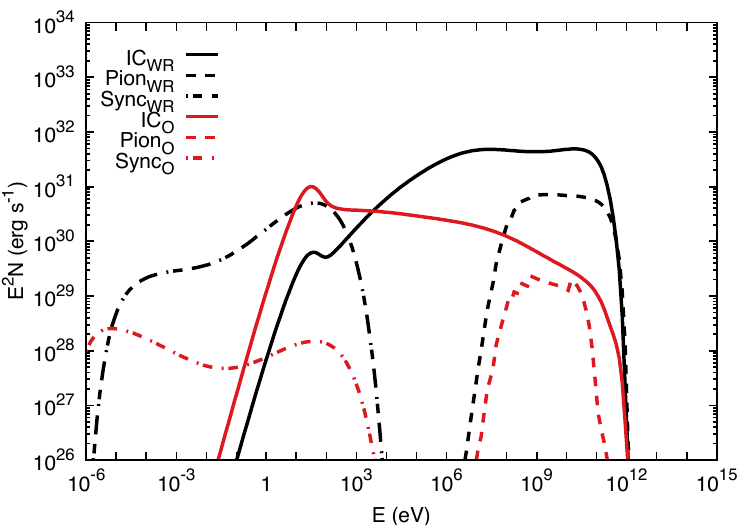}
\caption{The contribution of each shock to the intrinsic non-thermal emission
  from our standard model. The emission from particles accelerated by
  the WR-shock is shown by the black lines, while the emission from
  the O-shock accelerated particles is shown by the red lines. Except
  for a narrow range at $\sim 10-100$\,eV, the WR-shock completely dominates the
  non-thermal emission.}
\label{fig:stanCWB_ntEmission_WRandO}
\end{figure}

The spectral shape of the inverse Compton emission around
$1-10^{3}$\,eV is rather
unexpected. Fig.~\ref{fig:stanCWB_ntEmission_WRandO} shows that this
is due to emission from the particles accelerated at the O-shock, but
a ``bump''in the emission is also seen from particles accelerated at
the WR-shock. Tests show that in both cases this ``bump'' is produced
by electrons with Lorentz factors $\gamma < 2$. It arises due to the
cooling experienced by the downstream electrons, which gives rise to a
``peaked'' particle distribution \citep[see Figs.~6, 7 and 10
in][]{Pittard:2020}.  Removing the emission from these particles
creates a smooth downturn at these
energies. Fig.~\ref{fig:stanCWB_ntEmission_WRandO} also shows that the
total emission is dominated by particles accelerated at the WR-shock,
and that the emission from the O-shock is noticeably softer. This is
expected given the particle distributions shown in
Figs.~\ref{fig:stanCWB_mtmSpectra} and~\ref{fig:stanCWB_mtmSpectra2}.

\begin{figure*}
\includegraphics[width=8.0cm]{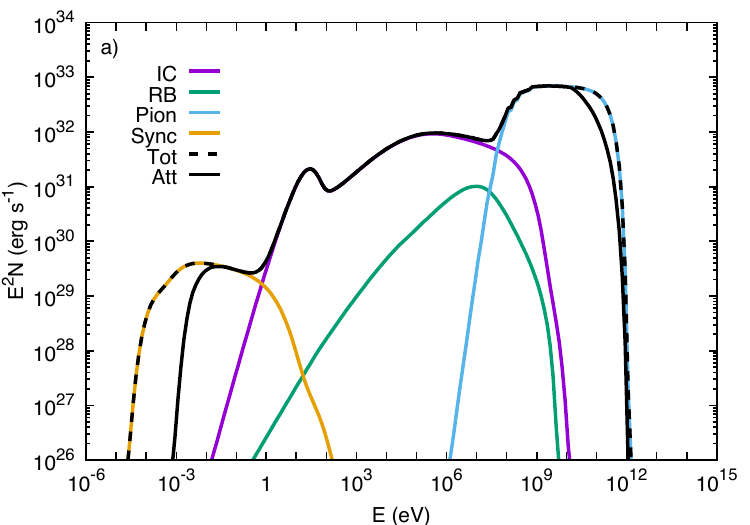}
\includegraphics[width=8.0cm]{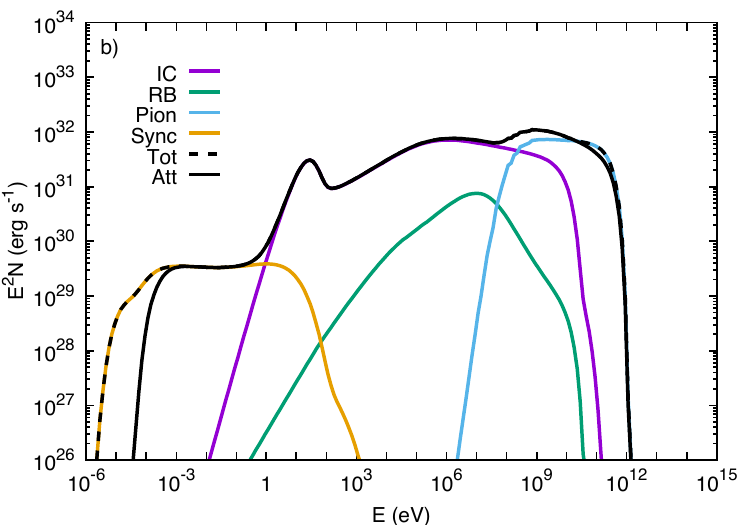}
\includegraphics[width=8.0cm]{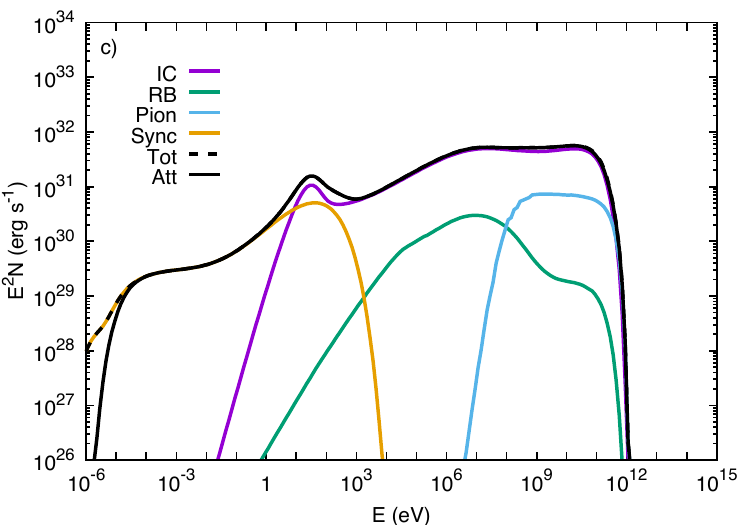}
\includegraphics[width=8.0cm]{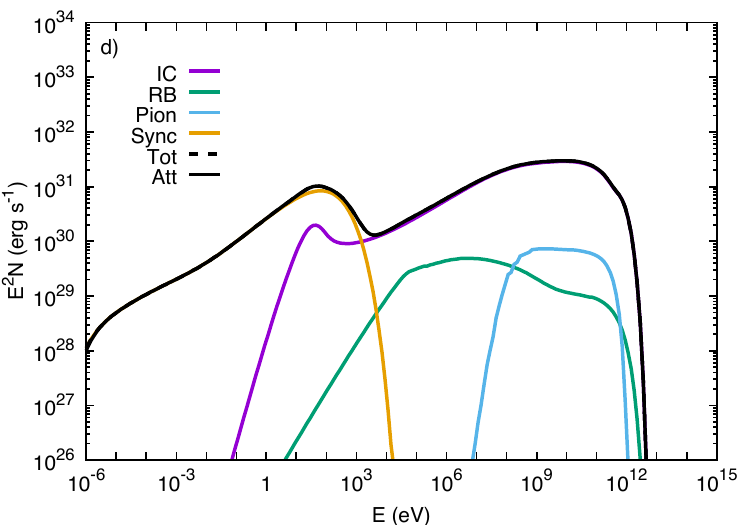}
\includegraphics[width=8.0cm]{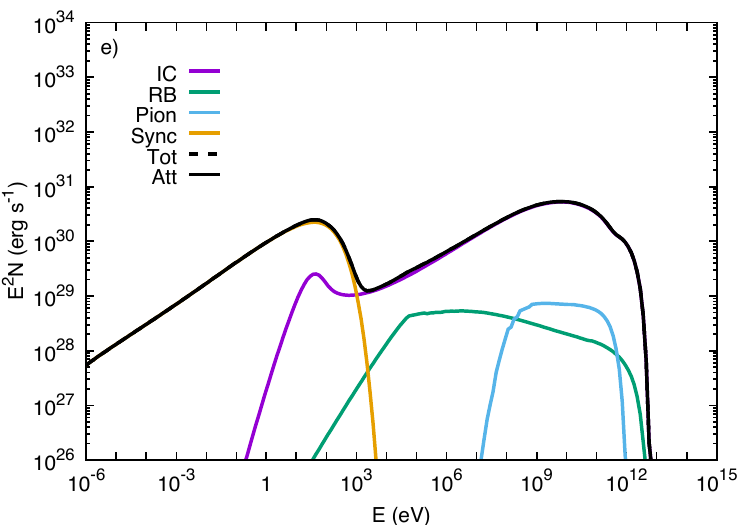}
\includegraphics[width=8.0cm]{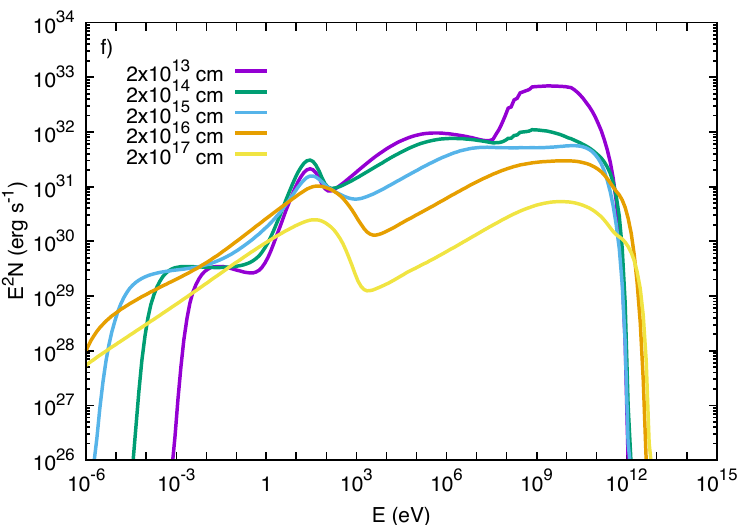}
\caption{The dependence of the non-thermal emission on the stellar
  separation, $D$, which varies between each panel: a)
  $D=2\times10^{13}\,{\rm cm}$; b) $D=2\times10^{14}\,{\rm cm}$; c)
  $D=2\times10^{15}\,{\rm cm}$; d) $D=2\times10^{16}\,{\rm cm}$; e)
  $D=2\times10^{17}\,{\rm cm}$. All other parameters are unchanged. In
  panel f) the total attenuated emission from each of these models is
  plotted. In all cases $\phi=90^{\circ}$.}
\label{fig:stanCWB_ntEmission_varDsep}
\end{figure*}

\subsection{The effect of varying the stellar separation}
Fig.~\ref{fig:stanCWB_ntEmission_varDsep} shows how the non-thermal
emission changes as the stellar separation is varied. It is clear that
the $\pi^{0}$-decay emission increases steadily as $D$ decreases
\citep[scaling as $D^{-1}$ - see][]{Pittard:2020}. So while the
non-thermal spectrum at large $D$ is dominated by synchrotron and
inverse Compton emission (at low and high energies, respectively), at
closer separations the high energy emission becomes dominated by the
$\gamma$-rays created by $\pi^{0}$-decay (emission from secondary
electrons may also become important - see
Sec.~\ref{sec:secondaryElectrons}). At $D=2\times10^{17}$\,cm the
synchrotron emission dominates up to $E=1$\,keV, while at higher
energies inverse Compton emission takes over. As $D$ decreases the
spectral shape of the synchrotron emission changes quite markedly, due
to a softening of the non-thermal electron spectrum. The maximum
energy of the inverse Compton emission is $\gtsimm 10^{12}$\,eV at
$D \gtsimm 10^{15}$\,cm, but decreases for closer separations, being
$\sim10^{10}$\,eV when $D\sim10^{13}$\,cm. $\gamma$-$\gamma$ absorption
only becomes significant at $D\ltsimm10^{14}$\,cm.

\citet{Pittard:2020} showed that the emission from non-thermal
electrons varies in a more complicated way with $D$. If the cooling
length of the non-thermal electrons is greater than or of order the
size of the WCR, then they fill the WCR and the emission also varies
as $D^{-1}$. However, if the non-thermal electrons cool more rapidly
then the emission will tend towards a constant value (i.e. be
independent of $D$). They also noted that as
$p_{\rm max,e} \propto D^{-1}$ (for an assumed scaling of
$B_{0} \propto D^{-1}$), this would drive further changes in the
emission with $D$.

Panel f) in Fig.~\ref{fig:stanCWB_ntEmission_varDsep} shows the total
attenuated non-thermal spectrum at each distance. For
$E\gtsimm10^{3}$\,eV the emission generally increases with decreasing
$D$, though depending upon the energy, the increase is not always
steady or even strictly monotonic. As $D$ drops further the emission
plateaus, as predicted by \citet{Pittard:2020}. Free-free absorption
by the clumpy stellar winds curtails the low-frequency synchrotron
emission as $D$ decreases, with the turnover frequency scaling as
$\nu\propto D^{-10/7}$ \citep{Dougherty:2003}. The Razin effect
produces a characteristic cut-off frequency that is given by
$\nu_{\rm R} = 20 n_{\rm e}/B$. Since in our standard model
$B\propto 1/D$ and $n_{\rm e} \propto 1/D^{2}$, the cut-off frequency
scales as $\nu_{\rm R} \propto D^{-1}$. This is responsible for the
turndown in the {\em intrinsic} synchrotron emission seen in
Fig.~\ref{fig:stanCWB_ntEmission_varDsep}.

\subsection{The effect of varying the stellar wind magnetic field}
The pre-shock magnetic field depends on the
strength of the magnetic field at the stellar surfaces, $B_{*}$, and
the rotation speed of the star, $v_{\rm rot}$. The latter affects how
tightly wound the field-lines are in the equatorial plane of the
star. In the extreme case that the stars are not rotating the stellar
wind drags the field lines into a radial configuration. In the
following we vary both $B_{*}$ (specifically the surface magnetic
field of the O-star, $B_{\rm *O}$) and $v_{\rm rot}$ to see how each may
change the particle acceleration and non-thermal emission.

\subsubsection{Changing the surface magnetic field}
\label{sec:surfaceB}
We first explore changing $B_{*}$. Since $B_{0}$ is higher for the
O-shock than it is for the WR-shock in the standard model (see
Fig.~\ref{fig:stanCWB_shockstats1}), we reduce the surface magnetic
field strength of the O-star to $B_{\rm *O}=10$\,G (the standard model
has $B_{*}=100$\,G for both stars). This results in an on-axis
pre-shock magnetic field strength of 2.1\,mG and an Alfv\'{e}nic Mach
number of 155 at the O-shock. The result is that the on-axis O-shock
becomes much more efficient at accelerating particles than before,
with 45\% of the incoming kinetic flux now turned into non-thermal
particles flowing downstream from the shock. This is a greater
efficiency than the on-axis WR-shock (which is at 23\%), and is also
manifest as a higher compression ratio for the O-shock
($R_{\rm tot}=7.2$) in this situation.

We find that the particle acceleration process behaves non-linearly
with the magnetic field strength at the shock. As the surface magnetic
field of the O-star reduces from $100$\,G the particle acceleration
efficiency at the O-shock first increases and then reduces again. This
is because of two competing effects. First, the acceleration
efficiency increases as the Alfv\'{e}nic Mach number of the shock
increases. Second, the maximum proton energy decreases
($E_{\rm max} \propto B_{0}$ - see Sec.~\ref{sec:pmax}) - this
eventually causes the acceleration to become inefficient.

\begin{figure*}
\includegraphics[width=17.5cm]{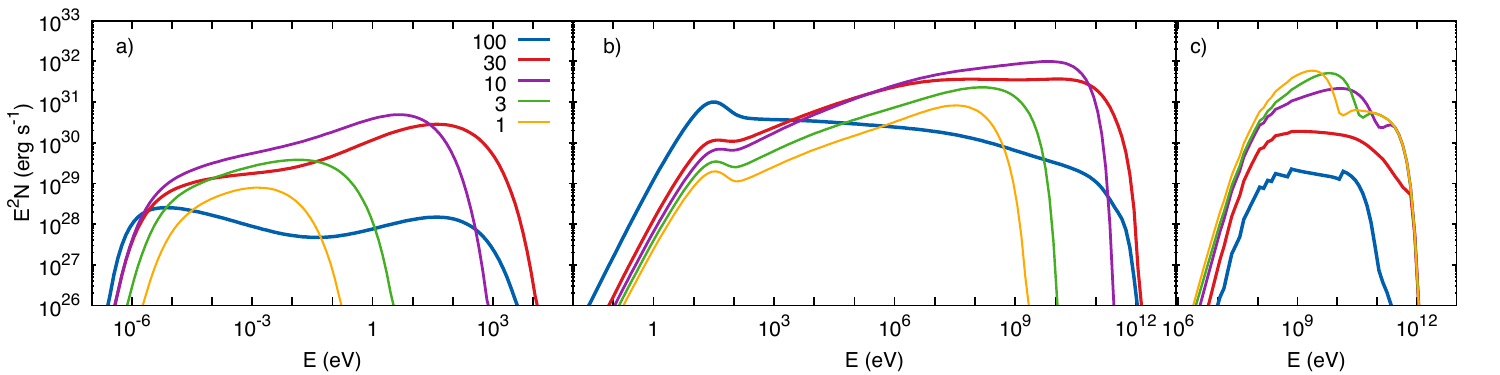}
\caption{The intrinsic non-thermal emission from the particles
  accelerated at the O-shock when the surface magnetic field of the
  O-star is varied. a) The synchrotron emission. b) The inverse
  Compton emission. c) The emission from $\pi^{0}$-decay.}
\label{fig:stanCWB_ntEmission_varBO}
\end{figure*}

This non-linear behaviour is manifest in the resulting non-thermal
emission which is shown in Fig.~\ref{fig:stanCWB_ntEmission_varBO}.
The dependence of the particle acceleration efficiency on
$M_{\rm A0x}$ and $p_{\rm max}$ results in the peak of the synchrotron
emission being obtained at an intermediate value of $B_{\rm *O}$.

Considering the IC emission in
Fig.~\ref{fig:stanCWB_ntEmission_varBO}b), we see that the IC emission
has a globally negative slope when $B_{\rm *O}=100$\,G, while the
curves for lower values of $B_{\rm *O}$ do not. This arises because of
the very steep particle distributions that are obtained when
$B_{\rm *O}=100$\,G (the standard case) due to the low values of
$S_{\rm tot}$ and $S_{\rm sub}$ (see
Fig.~\ref{fig:stanCWB_mtmSpectra}). As $B_{\rm *O}$ decreases and
$S_{\rm tot}$ and $S_{\rm sub}$ increase, the IC emission attains a
globally positive slope.

\subsubsection{Radial stellar magnetic fields}
\label{sec:radialB}
We now explore how the particle acceleration and emission changes if
we assume that the stars do not rotate. This results in a radial
magnetic field in each stellar wind, which declines as $r^{-2}$
(instead of a toroidal field that declines as $r^{-1}$). Hence this
change affects both the strength of the pre-shock magnetic field, and
its orientation to the shock. On the WCR axis the shocks become
parallel (compared to almost perpendicular in the standard model).

\begin{figure*}
\includegraphics[width=17.5cm]{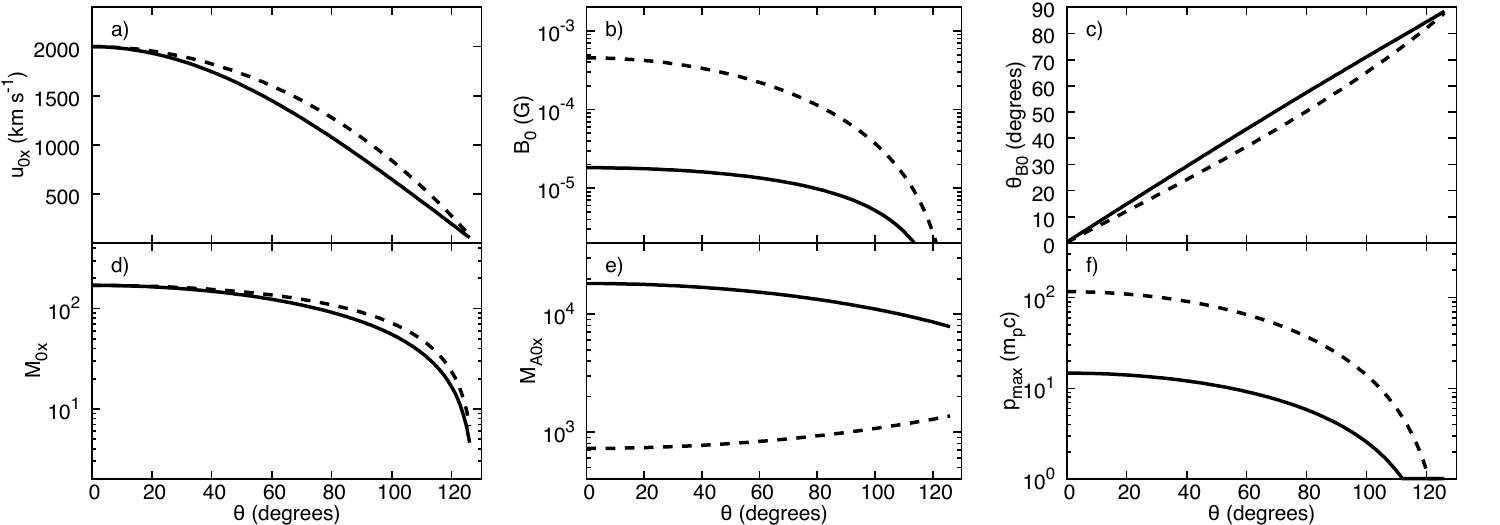}
\caption{As Fig.~\ref{fig:stanCWB_shockstats1} but for a model where
  the stars are not rotating, resulting in a radial magnetic field in
  each wind. All quantities are now independent of $\Phi$.}
\label{fig:stanCWBmodel7_shockstats1}
\end{figure*}

Fig.~\ref{fig:stanCWBmodel7_shockstats1} shows the pre-shock
quantities as a function of the angle $\theta$ from the secondary star
for the WR and O winds. Because the magnetic field in each stellar
wind is now radial, and drops as $r^{-2}$, the pre-shock magnetic flux
density is considerably lower than in the standard model, especially
for the WR-shock. This results in both shocks becoming highly
super-Alfv\'{e}nic ($M_{\rm A0x} > 10^{4}$ for the WR-shock, and
$M_{\rm A0x} \sim 10^{3}$ for the O-shock). Both shocks are parallel
on-axis and become nearly perpendicular far off-axis. The reduced
magnetic field strength also lowers the maximum momentum that the
non-thermal protons attain (again, particularly for the WR-shock).

\begin{figure*}
\includegraphics[width=17.5cm]{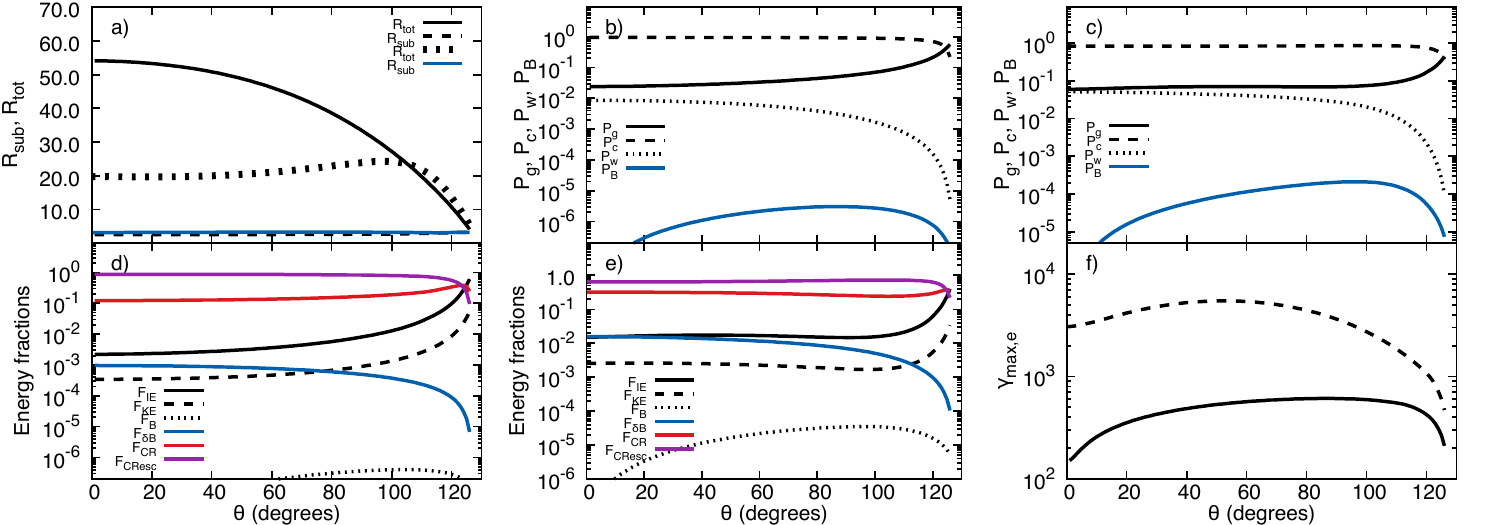}
\caption{As Fig.~\ref{fig:stanCWB_shockstats2} but for a model where
  the stars are not rotating, resulting in a radial magnetic field in
  each wind. All quantities are now independent of $\Phi$.}
\label{fig:stanCWBmodel7_shockstats2}
\end{figure*}

Fig.~\ref{fig:stanCWBmodel7_shockstats2} shows the post-shock
quantities as a function of the angle $\theta$ from the secondary
star. Both the WR-shock and O-shock are now extremely efficient
particle accelerators, and very high compression ratios are
obtained. The latter occurs despite creation of non-negligible
magnetic turbulence at the shock because of the very low magnetic
field strength upstream. On axis the WR-shock puts 12 and 87 per cent
of the incoming kinetic flux into non-thermal particles that flow
downstream and escape upstream, respectively. For the O-shock these
numbers are 32 and 65 per cent. The thermal X-ray emission from the
WCR (not calculated) will be much softer from this model than the
terminal speeds of the winds would suggest, because of the
significantly lower post-shock temperatures that are obtained (as a
large part of the input mechanical energy is used for particle
acceleration).  The turbulent magnetic field component dominates the
uniform field component, for both shocks and in all locations, by more
than an order of magnitude.

Another big change is the dramatic reduction in the maximum Lorentz
factor of the non-thermal electrons. We see that $\gamma_{\rm max,e}$ drops from
$\sim 10^{6}$ with a toroidal stellar magnetic field (see
Fig.~\ref{fig:stanCWB_shockstats2}f) to $\sim 10^{3}$ when the field
is radial. This is due to several factors: i) the large reduction in
the flow speed immediately prior to the subshock (due to the large
compression in the subshock in this model, $u_{1}=u_{0}R_{\rm
  sub}/R_{\rm tot} = 92\kmps$, compared to $1700\kmps$ in the standard
model); ii) the low value of the magnetic field immediately prior to
the subshock ($B_{1} = 2.6\times10^{-5}$\,G in this model, compared to
$5\times10^{-3}$\,G in the standard model); iii) the strongly turbulent
post-shock magnetic field ($B_{\rm 2,tot} = 0.044$ in this model,
versus 0.02 in the standard model). Factors i) and ii) strongly reduce
the acceleration rate of the electrons
(cf. Eq.~\ref{eq:dgammadtElectrons}), by about a factor of $10^{5}$,
while iii) increases the synchrotron loss rate by a factor of $\approx 5$.

\begin{figure}
\includegraphics[width=8.0cm]{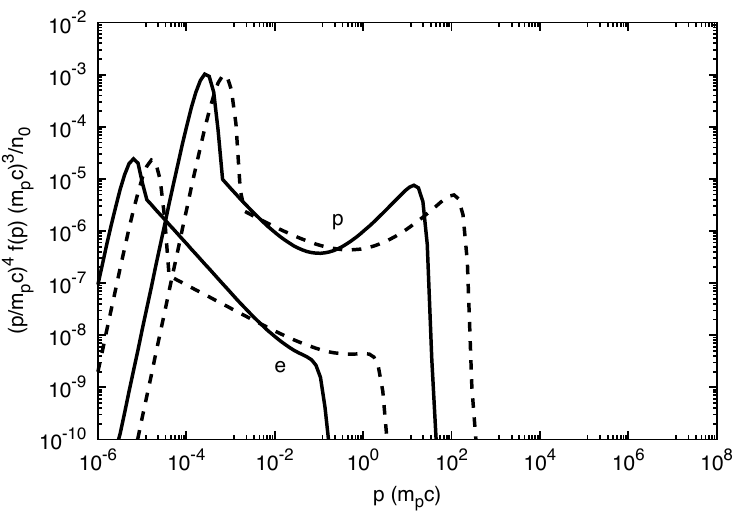}
\caption{The proton and electron distributions for the WR-shock (solid
  line) and O-shock (dashed line) for $\theta=0^{\circ}$ when the
  magnetic field in each stellar wind is radial. The distributions are
  independent of $\Phi$. For the WR-shock, $R_{\rm tot}
  = S_{\rm tot} = 54.1$ and $R_{\rm sub} = S_{\rm sub} = 2.5$. For the
  O-shock, $R_{\rm tot} = 19.8$ and $R_{\rm sub} = 3.09$, while
  $S_{\rm tot} = 19.6$ and $S_{\rm sub} = 3.06$. For both shocks
  $n_{0} = 1.3\times10^{5}\,{\rm cm^{-3}}$. The thermal peaks are
  visible at low momenta. $D = 2\times10^{15}$\,cm.}
\label{fig:stanCWBmodel7_mtmSpectra}
\end{figure}

Fig.~\ref{fig:stanCWBmodel7_mtmSpectra} shows the on-axis particle
distributions immediately downstream of the subshock. The strong
concave curvature to the distributions indicates the significant
modification of the shocks. The O-shock now contributes similarly to
the non-thermal particle population, whereas in the standard model the
O-shock contributed very little
(cf. Fig.~\ref{fig:stanCWB_mtmSpectra}). Neither shock accelerates
particles to particularly high energies, and as we have seen the
electron maximum energy is considerably reduced. The thermal peak
shows a significant shift to lower momenta, particularly for the
WR-shock, indicating the considerable reduction in post-shock
temperature.

\begin{figure}
\includegraphics[width=8.0cm]{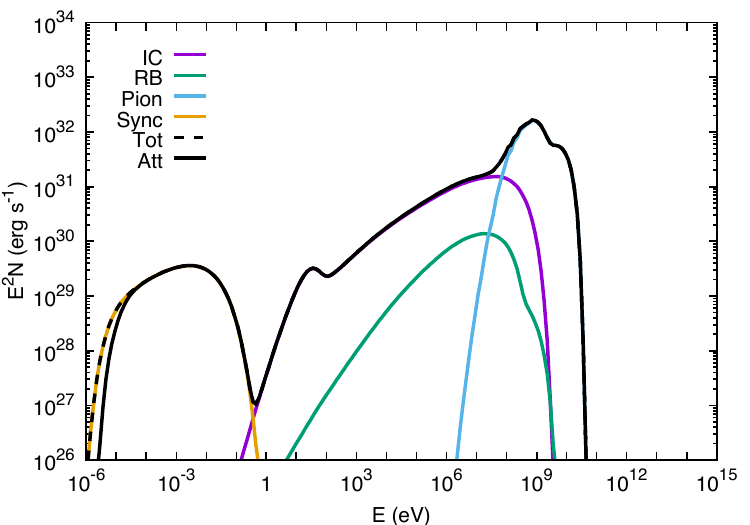}
\caption{The non-thermal emission when the stellar rotation speeds are
  zero (in the standard model $v_{\rm rot}/\vinfty=0.1$). $D = 2\times10^{15}$\,cm.}
\label{fig:stanCWB_ntEmission_vrot0}
\end{figure}

In Fig.~\ref{fig:stanCWB_ntEmission_vrot0} we show the non-thermal
emission from this model. The differences in the non-thermal particle
distributions compared to the standard model result in significant
differences to the non-thermal emission. First, we see a dramatic dip
in the emission between energies of $0.1-10$\,eV. This is caused by
the significantly lower energies attained by the non-thermal
electrons, which causes a reduction in the number of energy decades
that the synchrotron (and inverse Compton) emission extends
over. Second, the synchrotron and inverse Compton emission are both
significantly weaker. Third, the reduction in $p_{\rm max}$ also
lowers the maximum energy of the $\pi^{0}$-decay emission. Finally, we
see that the $\pi^{0}$-decay emission is significantly stronger
compared to the standard model at the same separation. This is because
of the much higher density of the post-shock gas due to the increased
compression of the shocks, plus the much lower flow speed of this gas,
which means that the ratio of the non-thermal proton cooling timescale
to the flow timescale has much reduced.

\subsection{Secondary electron creation and emission}
\label{sec:secondaryElectrons}
In some situations we might expect secondary electrons to make an
important contribution to the overall emission. Secondary electrons
can be created when non-thermal protons interact with either thermal
protons or with photons. The former case is expected to be dominant in
CWBs (see App.~\ref{sec:appPPvsPgam}). The emission from secondary
electrons has the potential to dominate that from the primary
electrons (those accelerated at the shocks) because the former
originate from the non-thermal protons, which carry the majority of
the energy that the non-thermal particles have \cite[see,
e.g.,][]{Orellana:2007}. It is also possible to create secondary
electrons with higher maximum energies than the primary electrons
(since the former is given by $0.05\,E_{\rm p,max}$, where
$E_{\rm p,max}$ is the maximum proton energy, while the latter depends
on inverse Compton and synchrotron losses during shock
acceleration). The secondary electrons have the same slope in their
particle distribution as the primary protons.

In order for secondary electrons to dominate the emission the
non-thermal protons must lose a significant fraction of their energy
through collisions with thermal protons. The inelastic proton-proton
cross-section is energy dependent, but we can take
$\sigma_{\rm pp}\approx30\,$mb as a good approximation \citep[see,
e.g., Fig.~A2 in][and also Eq.~\ref{eq:cross_section_pp} in this
work]{Vila:2012}. The cooling rate of a non-thermal proton is then
\begin{equation}
\frac{1}{t_{\rm pp}} \approx cn_{\rm p}K_{\rm pp}\sigma_{\rm pp},
\end{equation} 
where $n_{\rm p}$ is the thermal proton number density and $K_{\rm
  pp}\approx0.5$ is the total inelasticity of the interaction.

For cooling to be effective we require
$t_{\rm pp} \ltsimm t_{\rm dyn} = D/v_{\rm ps}$, where $v_{\rm ps}$ is
the postshock flow speed. With
$v_{\rm ps} \sim v_{\infty}/R_{\rm tot}$ and
$n_{\rm p} \sim R_{\rm tot}\Mdot/(4 \pi D^{2} v_{\infty} m_{\rm
  H})$, this gives
\begin{equation}
\label{eq:secondaryElectronsDsep}
D \ltsimm 10^{12}\frac{\Mdot_{-5}R_{\rm tot}^{2}}{v_{\infty,3}}\,{\rm cm},
\end{equation}
where $\Mdot_{-5} = \Mdot/(10^{-5} \Msolpyr)$ and
$v_{\infty,3}=v_{\infty}/(1000\kmps)$. For our standard model
$\Mdot_{-5} = 2$ and $v_{\infty,3}=2$ for the primary star, and
$R_{\rm tot} \sim 4$, so secondary electrons should become important
when $D_{\rm sep} \ltsimm 2\times10^{13}\,{\rm cm}$.

\begin{figure}
\includegraphics[width=8.0cm]{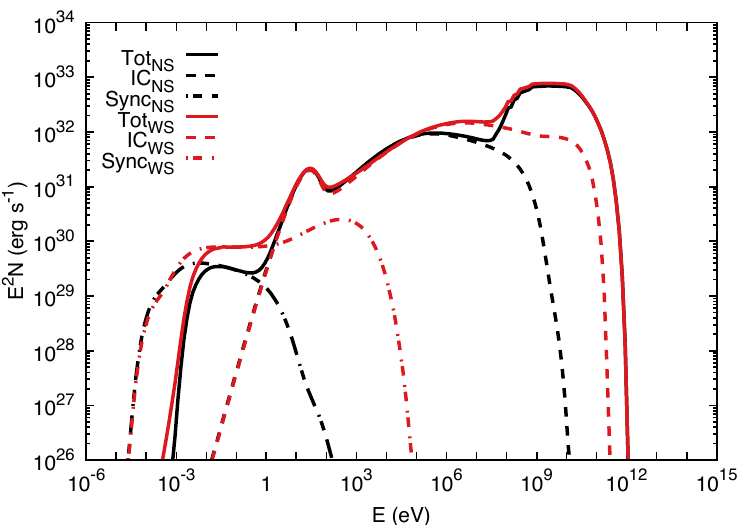}
\caption{Comparison of the intrinsic non-thermal emission from models
  with (subscript WS; red lines) and without (subscript NS; black
  lines) secondary electrons ($D=2\times10^{13}$\,cm). The secondary
  electrons boost the inverse Compton and synchrotron emission.}
\label{fig:stanCWB_ntEmission_secondaryElectrons}
\end{figure}

Fig.~\ref{fig:stanCWB_ntEmission_secondaryElectrons} compares the
leptonic and total non-thermal emission arising from models that
include or do not include secondary electrons. We see that the
secondary electrons give a significant boost to the high energy
inverse Compton emission (over the energy range
$E=10^{5}-10^{11}$\,eV) and synchrotron emission (over the energy
range $E=10^{-3}-10^{4}$\,eV), and indeed emit at higher energies than
the primary electrons are capable of. The emission produced by
secondary electrons is aided by the fact that they are continually
generated downstream (whereas the primary electrons cool as they flow
downstream and so only have a short opportunity to create the highest
energy synchrotron and inverse Compton emission).

Despite the significant boost to the leptonic emission that the
secondaries provide, however, there is relatively little change in the
spectrum of the total non-thermal emission. The reason is that at the
energies where this boost to the emission occurs, other processes tend
to be dominant (inverse Compton emission from the primary electrons
masks the synchrotron emission from secondary electrons at
$E\sim 10-10^{3}$\,eV, and $\pi^{0}$-decay emission at
$E\sim 10^{8}-10^{11}$\,eV masks the inverse Compton emission from
secondary electrons). Only between $10^{5}-10^{8}$\,eV do the
secondary electrons make a visible contribution to the total
emission\footnote{Note that the $\pi^{0}$-decay emission is the same
  in both models because cooling of the non-thermal protons is
  included in both - the difference is in whether the creation of
  secondary electrons is considered.}.

While secondary electrons do not appear to significantly affect the
total non-thermal emission for the standard model parameters with
$D \gtsimm 10^{13}$\,cm, they may be more important in systems with
higher stellar mass-loss rates and slower wind speeds, or if the
primary protons are able to interact with dense, radiatively cooled,
gas. \citet{White:2020} show that secondary electrons dominate the
emission between $E\approx 1-40$\,MeV in their ``off-periastron''
models of $\eta$~Carinae (see the top panel in their Fig.~3).

\begin{figure}
\includegraphics[width=8.0cm]{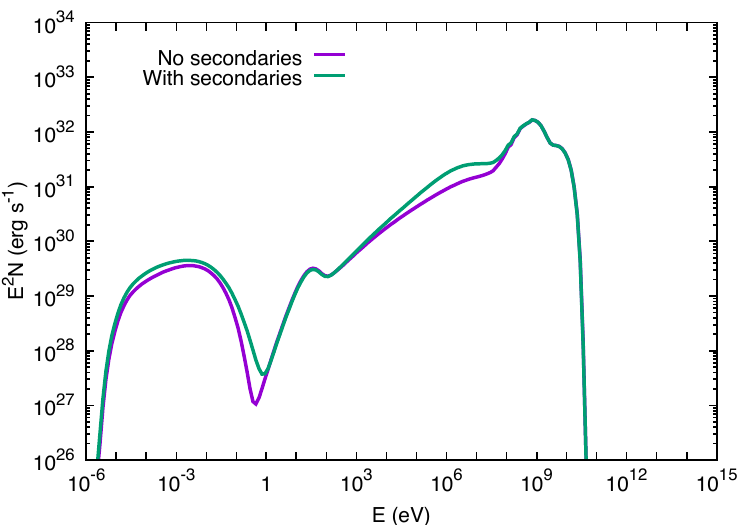}
\caption{Comparison of the non-thermal emission from models with and
  without secondary electrons. $D=2\times10^{15}$\,cm and $v_{\rm
    rot}/\vinfty=0.0$ (i.e. the stellar winds have radial magnetic fields).}
\label{fig:stanCWB_ntEmission_RadialBfieldsecondaryElectrons}
\end{figure}

Secondary electrons can also become important in situations where the
shocks are strongly modified and very high compression ratios are
achieved. Fig.~\ref{fig:stanCWB_ntEmission_RadialBfieldsecondaryElectrons}
compares the significance of secondary electrons in such models (see
Fig.~\ref{fig:stanCWBmodel7_shockstats2} for the $R_{\rm tot}$ values
in this case). While secondary electrons only become important for
stellar separations $D\ltsimm 2\times10^{13}\,$cm in models with the
standard parameters,
Fig.~\ref{fig:stanCWB_ntEmission_RadialBfieldsecondaryElectrons} shows
that secondaries can become important at much wider stellar
separations when the shocks are significantly modified. In this
particular case they are starting to add significantly to the emission
between $10^{5}-10^{7}$\,eV.

\section{Modelling the radio emission from WR\,146}
\label{sec:wr146}
Having explored how the particle acceleration and non-thermal emission
varies with stellar separation and the magnetic field in each wind,
and the conditions under which secondary electrons become important,
we now turn our attention to the modelling of a specific system.  We
choose WR\,146, a WC6+O8I-IIf system \citep{Lepine:2001}, because it
is amongst the brightest CWBs at radio wavelengths and is also one of
the few CWBs to be spatially resolved, with a southern thermal
component and a northern non-thermal component
\citep{Dougherty:1996,Dougherty:2000,O'Connor:2005}.  It has also been
resolved at optical wavelengths by HST \citep{Niemela:1998}, revealing
a projected stellar separation of $168\pm31$\,mas with the WR-star to
the south and the O-star to the north.  At 43\,GHz there is a
significant thermal contribution to the northern flux from the O-star
wind \citep{O'Connor:2005}. From the relative position of the
components, \citet{O'Connor:2005} inferred a wind momentum ratio of
$\eta=0.06\pm 0.15$. More recently, a search for polarized radio
emission has been made \citep{Hales:2017}.  WR\,146 is currently the
only CWB system to be detected at frequencies as low as 150\,MHz
\citep{Benaglia:2020}. The distance to WR\,146 is estimated as
$1.2\pm0.3$\,kpc \citep{Dougherty:1996}, which is compatible with the
{\em Gaia} DR2 estimate of $1.10^{+0.67}_{-0.36}$\,kpc
\citep{Rate:2020}. At a distance of $1.2$\,kpc, the projected stellar
separation is $2.9\times10^{15}$\,cm. Secondary electrons are not
expected to be important in this system (see
Eq.~\ref{eq:secondaryElectronsDsep}), and are therefore not included
in the following models.

In their X-ray analysis of WR\,146, \citet{Zhekov:2017} found that the
predicted theoretical X-ray flux from their models far exceeded the
observed emission. To bring the two measurements together required
either substantially reducing their adopted mass-loss rates (by a
factor of 10), or increasing the stellar separation $D$ (by a factor
of 66). The necessary change required for each variable in isolation
is rather implausible, which suggests that they need to vary in
combination, though even then the size of the required changes is
rather overwhelming. One then wonders what other process could be at
play. \citet{Zhekov:2017} note that models where the post-shock
electrons are not in temperature equilibration with the ions can
reduce the X-ray luminosity by another factor of two.
 
There seem to be three possible solutions to this problem. First, the
wind momentum ratio may be too high (\citet{Zhekov:2017} assumed that
$\eta=0.11$). A lower value would mean that a smaller fraction of the
WR wind is shocked, and since $L_{\rm x}\propto \eta$
\citep{Pittard:2018}, this would move the theoretical prediction
towards the observed flux. However, given the magnitude of the excess
emission this alone will not be enough. A second solution, which is
not incompatible with the previous one, is that a significant fraction of the kinetic
power of the stellar winds goes into non-thermal particles via
DSA. Both of these possibilities are investigated below. Finally, a
third possibility is that the post-shock flow is also not in
ionization equilibrium. This may impact the X-ray luminosity but a
detailed study is needed to determine at what level.

Our spectral models of WR\,146 are constrained by the observed flux from this
system. In the radio band we use the flux measurements by
\citet{Hales:2017} and \citet{Benaglia:2020}. We also include
measurements obtained using the VLA in combination with the VLBA Pie
Town antenna (see Table~\ref{tab:wr146vlaPlusPieTownData}). In the
X-ray band we use the on-axis ACIS-I {\em Chandra} pointed
observation (Obs ID 7426) taken on March\,$17^{\rm th}$\,2007 (PI
Pittard). This observation was designed to search for signs of weak
shock heating and shock modification. Finally, there
are also upper limits from 2 years of data from the {\em Fermi} satellite
\citep{Werner:2013}\footnote{\citet{Pshirkov:2016} do not detect
  WR\,146 in nearly 7 years of {\em Fermi} data, so should have been able to
  provide upper limits roughly $2\times$ lower. However, due to
  possible contamination from a
  complicated neighbourhood, they declined to provide upper limits.}. To
date, only one CWB has been detected at TeV energies
\citep[$\eta$\,Carinae;][]{HESS:2020}.

\begin{table}
\begin{center}
\caption[]{Flux and RMS measurements of WR\,146 obtained with the
  VLA in combination with the VLBA Pie Town antenna on October 1$^{\rm
    st}$ 2004 (the 22\,GHz data was obtained on November 8$^{\rm th}$ 2004). Where
two sources are resolved data is provided for both. At the lower
frequencies where this is not the case all the flux is assigned to the
northern source. From \citet{O'Connor:2005}.}
\label{tab:wr146vlaPlusPieTownData}
\begin{tabular}{lllll}
\hline
Frequency & N flux &  N RMS & S flux & S RMS \\   
(GHz)     & (mJy)  &  (mJy) & (mJy)  & (mJy) \\
\hline
1.465    & 71.92  & 1.4   & 0.0 & 0.0 \\
4.885    & 33.96  & 0.68  & 0.0 & 0.0 \\
8.435    & 23.46  & 0.47  & 0.0 & 0.0 \\
15.00    & 14.82  & 0.74  & 3.59 & 0.18 \\
22.46    & 10.33  & 0.52  & 5.17 & 0.26 \\
43.34    & 5.21   & 0.26  & 6.59 & 0.33 \\
\hline
\end{tabular}
\end{center}
\end{table}

\begin{table}
\begin{center}
  \caption[]{The parameters used in our final model of WR\,146. Our
    assumed $D=1.2\times10^{16}$\,cm implies that $i=76^{\circ}$ and
    $\phi=14^{\circ}$ (the O star is directly in front of the WR-star
    when $i=90^{\circ}$). The WR-star terminal wind speed is an
    average from \citet{Eenens:1994} and \citet{Willis:1997}. The
    O-star terminal wind speed is from the velocity ratio given in
    \citet{Dougherty:2000}. The O and WR-star luminosities are from
    O'Connor (private communication), and are estimated from
    \citet*{Vacca:1996} and the magnitude difference reported by
    \citet{Niemela:1998}. The hydrogen, helium and ``metal'' mass
    fractions are noted as $X$, $Y$ and $Z$, respectively. The WR
    abundances are from \citet{Nugis:2000}.}
\label{tab:wr146parameters}
\begin{tabular}{lll}
\hline
Parameter & WR-star & O-star\\
\hline
$\Mdot\,\,(\Msolpyr)$ & $2\times10^{-5}$ & $4\times10^{-6}$ \\
$v_{\rm \infty}\,\,(\kmps)$ & $2800$ & $1600$ \\
$L\,\,(\Lsol)$ & $2.3\times10^{5}$ & $7.9\times10^{5}$\\
$T_{\rm eff}\,\,$(K) & 49000 & 32000 \\
$R_{*}\,\,(\Rsol)$ & 6.6 & 28.9 \\
$X$ & 0.0 & 0.7381 \\
$Y$ & 0.744 & 0.2485 \\
$Z$ & 0.256 & 0.0134 \\
$B_{*}$\,\,(G) & 140 & 14 \\
$v_{\rm rot}/\vinfty$ & 0.1 & 0.1 \\
$f$ & 1.0 & 1.0 \\
\hline
\end{tabular}
\end{center}
\end{table}

\begin{table}
\begin{center}
\caption[]{The kinetic power of the winds, the kinetic flux at each
  shock, and the power put into non-thermal particles that are
  advected downstream or escape upstream from the shocks. All values
  are in $\ergps$.}
\label{tab:wr146modelOutputs}
\begin{tabular}{lccc}
\hline
Parameter & WR & O & Total\\
\hline
Wind kinetic power & $5.0\times10^{37}$ & $3.3\times10^{36}$ & $5.3\times10^{37}$\\
Input power at shock & $1.6\times10^{36}$ & $8.9\times10^{35}$ & $2.5\times10^{36}$\\
CR advection power & $3.4\times10^{35}$ & $2.4\times10^{35}$ & $5.7\times10^{35}$\\
CR escape power    & $6.7\times10^{34}$ & $6.9\times10^{34}$ & $1.4\times10^{35}$\\
\hline
\end{tabular}
\end{center}
\end{table}

\begin{figure*}
\includegraphics[width=8.0cm]{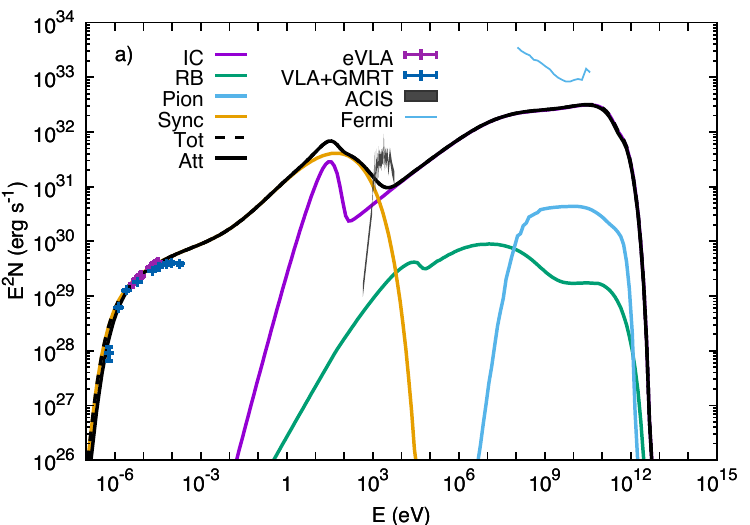}
\includegraphics[width=8.0cm]{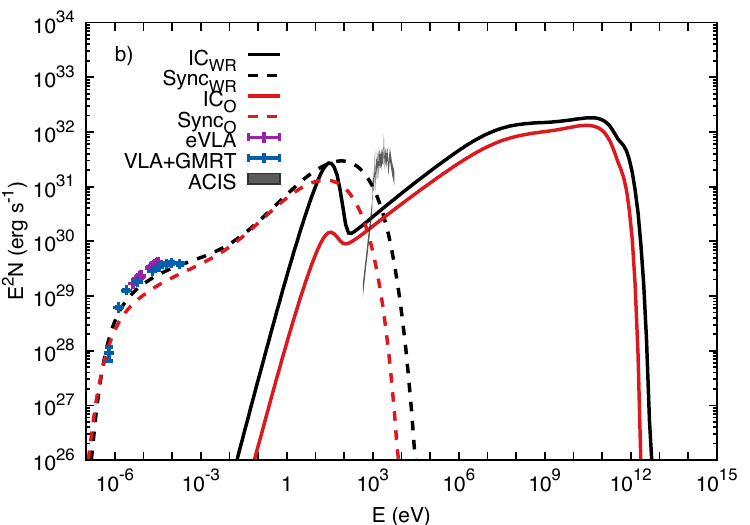}
\includegraphics[width=8.0cm]{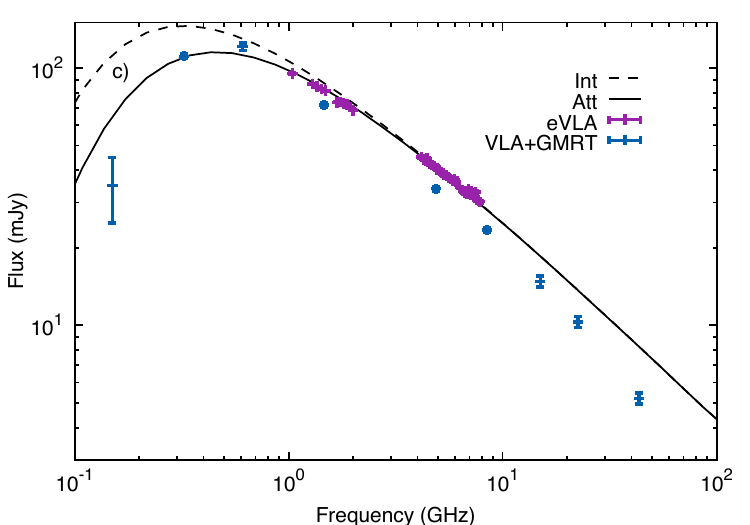}
\includegraphics[width=8.0cm]{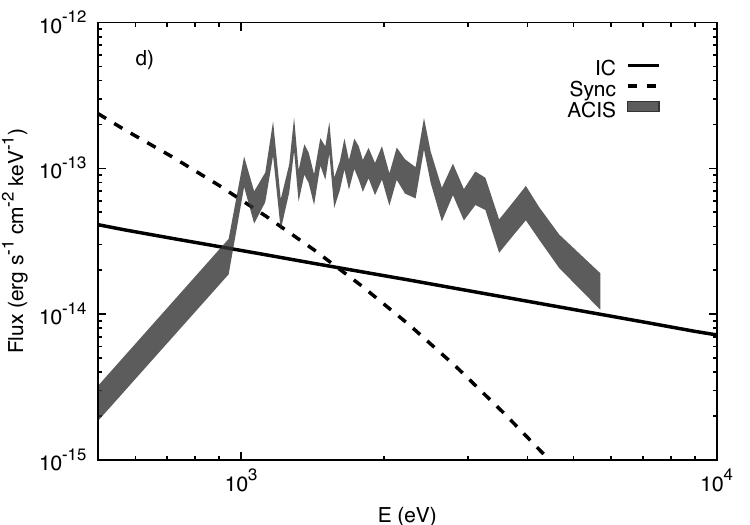}
\caption{Model spectra for WR\,146. a) The model inverse Compton,
  synchrotron, relativistic bremsstrahlung and $\pi^{0}$-decay
  emission are shown, together with the observed radio and thermal
  X-ray fluxes, and the upper limits from {\em Fermi}. b) The inverse
  Compton and synchrotron emission from particles accelerated at the
  WR-shock (black lines) and O-shock (red lines). c) The intrinsic and
  absorbed synchrotron radio emission from the model, and the observed
  radio emission. d) The observed X-ray emission and the non-thermal
  emission from the model. See Table~\ref{tab:wr146parameters} for the
  model parameters.}
\label{fig:wr146_model1}
\end{figure*}

\subsection{The modelling}
As it is unlikely that the stars are not rotating we adopt
$v_{\rm rot}/v_{\infty}=0.1$, which leads to a toroidal magnetic
field in each wind. We first attempted to fit the observational data
with the assumption that the system is viewed face-on
($D = 2.9\times10^{15}$\,cm; $i=0^{\circ}$; $\phi=90^{\circ}$). We
adopted somewhat lower mass-loss rates than usually found in the
literature, given the findings by \citet{Zhekov:2017}:
$\Mdot_{\rm WR}=2\times10^{-5}\,\Msolpyr$ and
$\Mdot_{\rm O}=2\times10^{-6}\,\Msolpyr$. With the observed terminal
wind speeds this gives a wind momentum ratio $\eta=0.057$. However, it
proved impossible to obtain a good match to the observed synchrotron
emission while simultaneously matching the turnover frequency at
$\nu\approx 450\,$MHz. In particular we found that the Razin turnover
frequency was always too high, and the synchrotron luminosity too
low. The former could be reduced by reducing the stellar surface
magnetic flux densities, but this led to lower synchrotron luminosity
(cf. Fig.~\ref{fig:stanCWB_ntEmission_varBO}). 

Since $\nu_{\rm R} \approx 20\,n_{\rm e}/B\,$, the Razin turnover
frequency can be lowered in the case that $B\propto1/D$ by increasing
$D$. Increasing the stellar separation to $D=1.2\times10^{16}$\,cm
($i = 76^{\circ}$; $\phi=14^{\circ}$) yielded $\nu_{\rm R}$ at the correct frequency, but
the synchrotron luminosity was still too low. 
To increase the synchrotron luminosity the O-star mass-loss rate was
increased to $\Mdot_{\rm O}=4\times10^{-6}\,\Msolpyr$, giving a wind
momentum ratio $\eta=0.11$. This increase in $\eta$ means that a
greater fraction of the WR-wind kinetic flux is intercepted by the
WCR. The kinetic flux of the O-wind also doubles. The increase in
$\Mdot_{\rm O}$ and $\eta$ does indeed produce stronger synchrotron
emission, and a reasonable match to the observational data is now
obtained (see Fig.~\ref{fig:wr146_model1}). With the assumed value of
$v_{\rm rot}/\vinfty=0.1$ we require $B_{\rm *WR}\approx140$\,G and
$B_{\rm *O}\approx14$\,G to match the turnover frequency and
synchrotron flux. The turndown below 1\,GHz is a combination of the
Razin effect and free-free absorption (see
Fig.~\ref{fig:wr146_model1}c). The latter is sensitive to the volume
filling factor of the clumps in the winds - here the winds are assumed
to be smooth (i.e. $f=1.0$; since the thermal free-free emission from
the stellar winds is not calculated in our model, $f$ only affects the
free-free absorption through the O-wind in the current model). While
our model is a good match to the recent eVLA data of
\citet{Hales:2017} and the GMRT data of \citet{Benaglia:2020}, it
matches less well the derived fluxes from the older VLA + Pie Town
data of \citet{O'Connor:2005}, which lie below the higher fluxes
reported by \citet{Hales:2017}. The parameters of our model are noted
in Table~\ref{tab:wr146parameters}.

In our model the non-thermal particles accelerated at the WR-shock
provide the majority of the emission, with the O-shock accelerated
particles typically contributing about a third to the total flux. The
WR-shock accelerated particles provide the highest energy inverse
Compton and synchrotron emission (see
Fig.~\ref{fig:wr146_model1}b). The non-thermal X-ray flux predicted by
the model is shown together with the observed X-ray emission in
Fig.~\ref{fig:wr146_model1}d). The inverse Compton emission barely
drops below the observed thermal emission at $E \approx 6$\,keV, while
the predicted synchrotron emission exceeds the observed thermal
emission at $E < 1$\,keV (note that no photoelectric absorption has
been applied to the model emission). In our model the synchrotron flux
at keV energies is sensitive to the value adopted for $\alpha$ in
Eq.~\ref{eq:alphaCut} and the assumption that the synchrotron loss
rate at the shock depends on $B_{\rm 2,tot}$ (this latter assumption
affects $p_{\rm max,e}$). Both of these ``close encounters'' with the
thermal X-ray emission may prove challenging to future models. In
theory, they may allow tight constraints to be placed on the O-star
luminosity (a higher luminosity would possibly decrease the maximum
energy of the non-thermal electrons and thus the maximum energy that
the synchrotron emission attains, but then would increase the
predicted inverse Compton emission, while a lower luminosity would
increase the maximum energy of the synchrotron emission). Future
models should also investigate whether radial magnetic fields in the
stellar winds produce a better match to the observations.

On axis the shocks put $\approx20$ per cent of the wind kinetic
flux into non-thermal particles, while a further 5 per cent goes into
non-thermal particles that escape upstream. Compression ratios of 4.7
are obtained. The upstream magnetic field strength is 0.72 and
0.93\,mG for the WR and O-shock respectively, while the post-shock
values are 3.4 and 4.3\,mG. 
 
Table~\ref{tab:wr146modelOutputs} notes the kinetic power of each wind,
the power available at each shock, and the power put into non-thermal
particles that are advected downstream or escape upstream of each
shock. The total power put into non-thermal particles is
$7.1\times10^{35}\,\ergps$, which represents an overall efficiency of
conversion of the power available at the shocks of 29 per cent. Just
over 1 per cent of the combined wind power of the stars goes into
non-thermal particles.

\subsection{Discussion}
Compared to the model in \citet{Zhekov:2017}, $\Mdot_{\rm WR}$ is
1.6 times lower, $D$ is 3.5 times higher, and $\eta$ is the
same. Since the thermal X-ray luminosity for an adiabatic system
scales as $L_{\rm x} \propto \Mdot^{2} \eta D^{-1}$
\citep{Stevens:1992,Pittard:2018}, our model should be 9 times
fainter by this measure. However, as 30 per cent of the available wind
power is put into cosmic rays rather than thermalised gas, it should
be $\approx 13$ times fainter overall. Unfortunately, this is still less
than the factor of $30-50$ reduction that \cite{Zhekov:2017} states is
required if $T_{\rm e} \leq T_{\rm ion}$. Perhaps non-equilibrium
ionization also has a role to play.

Turning our attention to the radio we note that although synchrotron
emission is intrinsically polarized, \citet{Hales:2017} found the
fractional linear polarization from the radio synchrotron emission
from WR\,146 to be less than 0.6 per cent. The lack of polarization is
naturally explained if the magnetic field is turbulent, and they
estimate that the field has a dominant random component with
$\delta B/B > 8$.  In contrast, we find that the emission weighted
value of $\delta B/B$ from our model is $\approx 2.4$ at frequencies
of $1-8$\,GHz. This suggests that some other process or mechanism may
be responsible for the lack of polarisation (see \citet{Hales:2017}
for a discussion of this). Alternatively, it may indicate that our
models of WR\,146 should have a magnetic field that is more
turbulent. This is achieved in our model with radial magnetic fields
in the stellar winds (see Sec.~\ref{sec:radialB}), where the turbulent
component dominates for both shocks in all locations by more than an
order of magnitude. The level of turbulence is interesting not least
because a high level of turbulence may lead to ultra-fast acceleration
in CWBs (and maximum energies above a few TeV), in contrast to
SNRs which appear to accelerate particles close to the Bohm limit \citep{Stage:2006}. 


While we have indicated the simple fitting that we have attempted, we
have certainly not exhausted all possibilities, and it is quite likely
that fits as least as good will be found with other model
parameters. This is because various trade-offs exist between the model
parameters. For instance, increasing $D$ generally leads to a drop in
emission, but this can be offset by increasing $\Mdot$. In addition,
$B_{*}$ and $v_{\rm rot}/\vinfty$ can be directly played off against
each other. Having said this, the model does place some
constraints. Too high values for the magnetic flux density result in
particle distributions that are too steep. Too low values for $B$
result in no or very weak acceleration, and/or too low values of
$p_{\rm max}$ and $p_{\rm max,e}$.  A more detailed investigation,
that will also model the free-free radio emission, the thermal X-ray
emission, and produce radio images, is left to future work.


\section{Summary and conclusions}
\label{sec:summary}
We report on the first particle acceleration model of colliding wind
binaries that applies a non-linear diffusive shock acceleration model,
with magnetic field amplification and relative motion of the
scattering centres, to oblique shocks. We find that:
\begin{enumerate}
\item The relative motion of the scattering centres with respect to the
fluid can be significant. When this occurs we obtain steeper
non-thermal particle distributions.
\item The particle acceleration is strongly dependent on the pre-shock
  magnetic field, and its efficiency can vary strongly along and
  between each shock. 
\item The particle acceleration efficiency and non-thermal emission
  can behave non-linearly with the magnetic field strength at the
  shock. As the pre-shock magnetic flux density decreases, an increase
  in acceleration efficiency due to the increasing Alfv\'{e}nic Mach
  number competes against a reduction in the maximum energy of the
  accelerated particles. This can result in the non-thermal emission
  peaking at intermediate values of the magnetic field strength.
\item The strength and angle to the shock normal of the pre-shock
  magnetic field depends strongly on whether the stellar winds have a
  toroidal field (i.e. the stars are rotating) or a radial field
  (i.e. the stars are non-rotating).
\item The non-thermal emission may be dominated by particles accelerated by
  one or the other shock, or may be roughly equally split between both
  shocks.
\item The shock precursors are typically smaller than the scale of the
  WCR.
\item Downstream of the shock the dominant pressure may be from the
  gas or from the cosmic rays.
\item In some locations along the shocks we find that $\delta B > B$,
  while the opposite is true in other locations. In our standard model
  we find that the WR-shock is largely turbulent while the O-shock is
  not. Whether or not a shock is turbulent depends sensitively on the
  model parameters, such as the strength of the surface magnetic field
  and rotation speed of the star. In some systems the synchrotron
  emission should not be significantly polarized, while in others it
  may be.
\item Local particle acceleration efficiencies for the downstream flowing
  cosmic rays of up to 30 per cent are obtained. Such values can arise
  when the shocks are perpendicular, oblique, or parallel. When the
  magnetic field in the stellar wind is radial, the lower pre-shock
  magnetic flux densities that result mean that up to nearly 90 per
  cent of the local kinetic flux may go into cosmic rays that escape
  upstream. Under other conditions the advected and escape cosmic ray
  energy fractions may be much reduced.
\item The gas compression ranges from $\leq 4$ to over 20 in some
  cases. High ratios have a significant effect on the strength of the
  emission from $\pi^{0}$-decay and secondary electrons, and will also
  affect the postshock temperature and the thermal X-ray emission.
\end{enumerate}

Given the large variation in the spectral indices of the non-thermal
particles seen in our models, it is clearly necessary to go beyond the
assumption that $f(p) \propto p^{-4}$ (equivalent to
$N(E) \propto E^{-2}$). While previous works have varied the spectral
index of the non-thermal particles as a model {\em input}
\citep[e.g.,][]{Pittard:2006a,Pittard:2006b,delPalacio:2016,delPalacio:2020},
our new model produces the spectral index as an {\em output}, and
allows it to vary along and between the shocks, and as a function of
energy or momentum. We draw attention to the fact that the values of
the energy index output from our standard model corresponds precisely to
the indices adopted by \citet{delPalacio:2016,delPalacio:2020} to
match the observed emission from HD\,93129A.

We also derive an analytical expression to determine when emission
from secondary electrons is expected to make an important contribution
to the total emission (Eq.~\ref{eq:secondaryElectronsDsep}). Such
secondaries can produce emission at higher energy than the primary
electrons, but we also show how the additional emission can sometimes
be masked by other emission processes.

Our new model has been applied to WR\,146, one of the brightest CWB
systems in the radio band. We are able to obtain a good match to the
radio flux, reproducing both the curvature of the eVLA data plus the
low frequency turnover. Our model is also consistent with other data:
the non-thermal emission is fainter than the observed thermal X-ray
emission and the {\em Fermi} upperlimit. The model converts
$\approx 30$ per cent of the kinetic wind power at the shocks into
non-thermal particles. If this WR+O system has a lifetime of
$\approx 3\times10^{5}$\,yr, it will put nearly $10^{49}$\,erg into
non-thermal particles during this evolutionary phase of the
stars. Significant energy may also go into cosmic rays during the
prior O+O phase which involves weaker winds but is longer lasting.

\section*{Acknowledgements}
We thank the referee for his/her encouragement and some useful suggestions. 
The calculations herein were performed on the DiRAC 1 Facility at
Leeds jointly funded by STFC, the Large Facilities Capital Fund of BIS
and the University of Leeds and on other facilities at the University
of Leeds. JMP was supported by grant ST/P00041X/1 (STFC, UK). GER was supported by grants PIP 0338 (CONICET, Argentina), PICT 2017-2865 (ANPCyT, Argentina) and the Spanish Ministerio de Ciencia e Innovaci\'{o}n (MICINN) under grant PID2019-105510GBC31 and through the ``Center of Excellence Mar\'{i}a de Maeztu 2020-2023'' award to the ICCUB (CEX2019-000918-M).

\section*{Data Availability}
The data underlying this article are available in the Research
Data Leed Repository, at \url{https://doi.org/XXX}.








\appendix

\section{Semianalytical Nonlinear Calculation of Particle
  Acceleration}
\label{sec:appDSA}
In this appendix we provide equations and the method for obtaining an
exact solution for the spatial and momentum distribution of particles
accelerated at a shock. The non-thermal particles generate Alfv\'{e}n
waves, and the magnetic turbulence and the cosmic rays dynamically
react back on the shock. The method is based on a 1D kinetic treatment
of parallel shocks developed by \citet{Amato:2005,Amato:2006} and
\citet{Caprioli:2009}, and modified by \citet{Grimaldo:2019} to
include a pressure term from a transverse component of the background
magnetic field. Like \citet{Grimaldo:2019} we do not consider how the
DSA efficiency changes with the obliquity of the shock - this
possibility is discussed further in Sec.~\ref{sec:obliquity}.  We
assume that all quantities change locally only in the $x$-direction
which is perpendicular to the shock and that the magnetic field lies
in the $x$-$z$ plane. Unlike \citet{Grimaldo:2019}, we assume that the
scattering centres move relative to the fluid at the Alfv\'{e}n
velocity.

The solution is obtained by iteratively solving the
diffusion-advection equation for the shock-accelerated particles. The
cosmic rays are described by their distribution function in phase
space $f(x,p)$ where $p$ is the particle momentum. Keeping only the
isotropic part, the diffusion-advection equation for a 1D
non-relativistic shock is:
\begin{multline}
\label{eq:diffusion-advection}
[u_{x}(x)-v_{\rm A}(x)]\frac{\partial f(x,p)}{\partial x} =
\frac{\partial}{\partial x}\left[D(x,p)\frac{\partial}{\partial
    x}f(x,p)\right] \\ + \frac{d [u_{x}(x)-v_{\rm
    A}(x)]}{dx}\frac{p}{3}\frac{\partial f(x,p)}{\partial p} + Q(x,p),
\end{multline}
where $u_{x}(x)$ is the flow speed in the x-direction, $v_{\rm A}(x)$ is the Alfv\'{e}n
velocity, $D(x,p)$ is the diffusion coefficient, and $Q(x,p)$
describes the injection of particles into the acceleration
process. The calculation is performed in the frame of the sub-shock
which is located at $x=0$; upstream is at $x < 0$ and downstream is at
$x>0$. In the following, quantities
evaluated far upstream are given the prefix ``0'', quantities
evaluated immediately upstream of the subshock are given the prefix
``1'', and quantities evaluated immediately downstream of the subshock
are given the prefix ``2''. 

The input values are the conditions far upstream: the flow density,
$\rho_{0}$, the flow velocity in the $x$-direction, $u_{0x}$, the gas
thermal pressure, $p_{\rm g0}$, the flux density of the unperturbed
magnetic field, $B_{0}$, and the angle it makes to the shock normal,
$\theta_{B0}$. In the following, velocities and pressures indicated
with capital letters are normalized by $u_{0x}$ and
$\rho_{0}u_{0x}^{2}$, respectively.

We consider only Alfv\'{e}n waves generated by the resonant-streaming
instability. The local Alfv\'{e}n velocity is given by
\begin{equation}
\label{eq:vA}
v_{\rm A}(x) = \frac{B(x)}{\sqrt{4 \pi \rho(x)}},
\end{equation}
where $\rho(x)$ and $B(x)$ are the local gas density and magnetic flux
density of the uniform field, respectively. $B_{x}$ is constant
throughout the shock, while $B_{z}$ changes in the shock precursor and
across the subshock. Hence $\theta_{B}$ is also a function of $x$.

We suppose that the diffusion coefficient is given by Bohm
diffusion\footnote{In Bohm diffusion the mean
  free path $l=r_{\rm L}$, where $r_{\rm L}=pc/(eB)$ is the Larmor
  radius. This results in a diffusion coefficient $D = r_{\rm L}c/3$, which is reasonable for
  situations with $\delta B/B \sim 1$. However, when the turbulence is
  very strong, the particles experience very strong scattering and the
  mean-free path becomes $l=r_{\rm L}B/\delta B$ \citep{Hussein:2014}. This
  results in a smaller diffusion coefficient and shorter acceleration
  timescales for the particles in the context of DSA. The maximum
  energy of particles may then be underestimated in the case of
  WR\,146, where the absence of polarization suggests very strong turbulence.}
in the self-generated magnetic field \citep[e.g.,][]{Jones:1991} so that
\begin{eqnarray}
D(x,p) & = & D_{\parallel}(x,p)\cos^{2}\theta_{B}(x) +
  D_{\perp}(x,p)\sin^{2}\theta_{B}(x), \nonumber \\
D_{\parallel}(x,p) & = & \frac{pc}{eB(x)}\frac{c}{3}, \nonumber\\
D_{\perp}(x,p) & = & \frac{1}{4}\frac{pc}{eB(x)}\frac{c}{3}.
\end{eqnarray}
$D_{\parallel}$ and $D_{\perp}$ are the diffusion coefficients
parallel and perpendicular to the magnetic field lines, respectively.

We assume that particle injection occurs only at the shock and
for particles with momentum $p_{\rm inj}$, such that
\begin{equation}
\label{eq:Qxp}
Q(x,p) = \eta \frac{\rho_{1}u_{1x}}{4\pi m_{\rm p} p_{\rm inj}^{2}}
\delta(p-p_{\rm inj})\delta(x),
\end{equation}
where we adopt the recipe of \citet*{Blasi:2005} for the fraction
$\eta$ of injected particles:
\begin{equation}
\label{eq:eta}
\eta = \frac{4}{3\sqrt{\pi}}(S_{\rm sub}-1)\chi^{3}e^{-\chi^{2}}.
\end{equation}
This prescription assumes that only particles with momentum $p_{\rm
  inj}\geq \chi p_{\rm th,2}$ can be accelerated, where $p_{\rm th,2}$
is the momentum of the thermal peak in the post shock gas. We follow
\citet{Caprioli:2009} and set $\chi=3.75$ in all of our simulations.

A very good approximation for the solution of
Eq.~\ref{eq:diffusion-advection} is \citep{Amato:2005,Amato:2006,Blasi:2007}
\begin{equation}
f(x,p) = f_{1}(p){\rm exp}\left[-\frac{(S_{\rm sub}-1)}{S_{\rm
      sub}}\frac{q(p)u_{0}}{3}\int^{0}_{-\infty}\frac{U_{x}(x') - V_{\rm A}(x')}{D(x',p)}dx'\right],
\end{equation}
where $V_{\rm A}(x) = v_{\rm A}(x)/u_{0x}$, $f_{1}=f(0,p)$ and $q(p) =
-\frac{{\rm d\,log}\,f_{1}(p)}{{\rm d\,log}\,p}$ is the spectral slope
at the shock location.

\citet{Blasi:2002} showed that $f_{1}(p)$ can be written as
\begin{multline}
\label{eq:f1p}
f_{1}(p) = \left(\frac{3 S_{\rm tot}}{S_{\rm tot}U_{px}(p)-1}\right)
\frac{\eta \rho_{1}}{4 \pi m_{\rm p} p_{\rm inj}^{3}} \\ \times{\rm
  exp}\left[-\int^{p}_{p_{\rm inj}} \frac{3 S_{\rm tot} U_{px}(p')}{S_{\rm tot}U_{px}(p')-1}\frac{dp'}{p'}\right],
\end{multline}
where
\begin{equation}
\label{eq:Up}
U_{px}(p) = U_{1x} - V_{\rm A1} -
\frac{1}{f_{1}(p)}\int^{0}_{-\infty}f(x,p)\frac{{\rm d}[U_{x}(x)-V_{\rm
    A}(x)]}{\rm dx} {\rm dx}.
\end{equation}
$u_{x}(p) = U_{px}(p)u_{0x}$ is the mean velocity effectively felt by a
particle with momentum $p$ in the upstream region.

The method of solution is based also on the momentum flux conservation
equation, normalized to $\rho_{0}u_{0x}^{2}$:
\begin{equation}
\label{eq:mtmconservation}
U_{x}(x) + P_{\rm c}(x) + P_{\rm w}(x) + P_{\rm g}(x) + P_{\rm B}(x) =
1 + P_{\rm g0} + P_{\rm B0},
\end{equation}
where the normalized thermal pressure $P_{\rm g0} = 1/(\gamma M_{\rm 0x}^{2})$ and $M_{\rm
  0x}^{2} = \rho_{0}u_{0x}^{2}/(\gamma p_{\rm g0})$. The
normalized pressure in cosmic rays is
\begin{equation}
\label{eq:Pcx}
P_{\rm c}(x) = \frac{4\pi}{3\rho_{0}u_{0x}^{2}} \int_{p_{\rm
    inj}}^{\infty} p^{3}v(p)f(x,p)\,{\rm d}p,
\end{equation}
where $v(p)$ is the velocity of a particle with momentum $p$, 
while the normalized pressure in magnetic turbulence generated via the
resonant streaming instability is
\begin{equation}
\label{eq:Pwx}
P_{\rm w}(x) = \frac{v_{\rm A}(x)}{4u_{0x}}\frac{1 -
  U(x)^{2}}{U(x)^{3/2}} \cos\theta_{B0}.
\end{equation}
If the heating of the background gas, with adiabatic index
$\gamma$, in the precursor is purely adiabatic, the normalized gas
pressure is
\begin{equation}
\label{eq:Pgx}
P_{\rm g}(x) = \frac{U_{x}(x)^{-\gamma}}{\gamma M_{0x}^{2}}.
\end{equation}
The $z$-component of the magnetic field exerts a normalized pressure
\begin{equation}
\label{eq:PBx}
P_{B}(x) = \frac{B_{z}(x)^{2}}{8\pi \rho_{0} u_{0x}^{2}},
\end{equation}
where
\begin{equation}
\label{eq:Bz}
B_{z}(x) = \left(\frac{M_{A0x}^{2} -
    \cos^{2}\theta_{B0}}{U_{x}(x)M_{A0x}^{2} - \cos^{2}\theta_{B0}}\right)B_{0z}
\end{equation}
and $M_{A0x} = u_{0x}/v_{A}$. If the shock is not strictly parallel $P_{\rm B}$ is present.

Given a value for $U_{1x} = R_{\rm sub}/R_{\rm tot}$, the normalized
values $P_{\rm g1}$, $P_{\rm w1}$, $P_{\rm c1}$, and $P_{\rm B1}$ are
determined. $R_{\rm sub}$ can then be determined by solving the
third-order equation
\citep[cf.][]{Decker:1988,Vainio:1999,Grimaldo:2019}\footnote{Note
  that \citet{Vainio:1999} use an incorrect definition for $\beta$,
  which misses out a factor of $2/\gamma$. In addition, Eq.\,16 in
  \citet{Caprioli:2009} for $R_{\rm sub}$ is incorrect.}
\begin{equation}
a_{3}R_{\rm sub}^{3} + a_{2}R_{\rm sub}^{2} + a_{1}R_{\rm sub} + a_{0}
= 0,
\end{equation}
with coefficients
\begin{eqnarray}
a_{3} & = & [(\gamma-1)(1+\lambda)M_{A1x}^{2} +
            \gamma\beta_{1}\cos^{2}\theta_{B1}]\cos^{2}\theta_{B1},\nonumber\\
a_{2} & = & \biggl[[2(1+\lambda) - \gamma(1 + \cos^{2}\theta_{B1} +
            \lambda)]M_{A1x}^{2} \nonumber \\
 & &  -[1 + \lambda + \gamma(2\beta_{1} + 1 +
    \lambda)]\cos^{2}\theta_{B1}\biggr]M_{A1x}^{2},\nonumber\\
a_{1} & = & [(\gamma-1)M_{A1x}^{2} + \gamma(1+\lambda +
            \cos^{2}\theta_{B1} + \beta_{1})\\
 & & + 2\cos^{2}\theta_{B1}]M_{A1x}^{4},\nonumber\\
a_{0} & = & -(\gamma+1) M_{A1x}^{6}\nonumber.
\end{eqnarray}
In the above equations, $\lambda = (\delta B_{1}/B_{1})^{2}$, $M_{A1x}
= u_{1x}/v_{A}$, and $\beta$ is the plamsa beta given by $\beta = 8\pi
p_{\rm g}/B^{2}$. This then yields a value for $R_{\rm tot}$. 

To solve these equations a numerical grid spanning $x/x_{*} =
10^{-10}-100$, where $x_{*}=-D(p_{\rm max})/u_{0x}$ and $D(p_{\rm
  max})$ is calculated using $p_{\rm max}$, $B_{1x}$ and $B_{1z}$. A momentum grid spanning
$p_{\rm inj}$ to $p_{\rm max}$ is also used. Each grid has 600 bins
distributed logarithmically. The following solution method is used:

\begin{enumerate}
\item For a given $p_{\rm max}$, and upstream flow parameters
  $\rho_{0}$, $u_{0x}$, $T_{0}$, $B_{0}$ and $\theta_{B0}$, guess an initial value of
  $U_{1x}$. Given $U_{1x}$, calculate $P_{\rm g1}$, $P_{\rm w1}$, $P_{\rm
    c1}$, $B_{1z}$ and $P_{\rm B1}$. 
\item Set $U_{x}(x) = U_{1x}$ for all $x$ and $U_{px}(p) = U_{1x}$ for
  all $p$ on the momentum grid. Set $P_{\rm g}(x) = P_{\rm g1}$, $P_{\rm B}(x) = P_{\rm B1}$,
  $P_{\rm w}(x) = P_{\rm w1}$ and $P_{\rm c}(x) = P_{\rm c1}$ for all
  $x$. Set $B_{z}(x) = B_{1z}$ for all $x$. Determine all immediate pre-shock quantities (with subscript 1).
\item Determine $R_{\rm sub}$, $R_{\rm
    tot}$, all immediate post-shock quantities (with subscript 2), $S_{\rm
    sub}$, $S_{\rm tot}$, $p_{\rm inj}$ and $\eta$. $B_{2x} = B_{1x}$
  and $B_{2z} = R_{\rm sub} B_{1z}$.
\item Calculate $\rho(x)$, $B(x)$ and $V_{\rm A}(x)$ for all $x$. The magnetic
  flux density of the uniform field is given by $B(x) = \sqrt{B_{x}^{2}(x) + B_{z}^{2}(x)}$.
\item Calculate $f_{1}(p)$.
\item Calculate $P_{\rm c}$ from $f_{1}(p)$ and compare to $P_{\rm
    c1}$. Let $K = P_{\rm c1}/P_{\rm c}$. Renormalize $f_{1}(p)$ by
  multiplying by $K$.
\item If $K$ is converged with its previous value goto item (xiii)
  below. Otherwise calculate $q(p)$, $D(x,p)$, $f(x,p)$, and $P_{\rm
    c}(x)$.
\item Calculate $U_{x}(x)$ from Eq.~\ref{eq:mtmconservation}. To achieve faster convergence average
    the flow profile $U_{x}(x)$ with its previous value.
\item Update $\rho(x)$, $P_{\rm g}(x)$, $P_{\rm w}(x)$, $B_{z}(x)$,
  $B(x)$, $P_{\rm B}(x)$, and $V_{\rm A}(x)$.    
\item Update the immediate pre-shock quantities from these values (e.g.,
  $U_{1x}=U_{x}(x=0^{-})$ where $x=0^{-}$ is immediately upstream of
  the subshock).
\item Determine $R_{\rm sub}$, $R_{\rm
    tot}$, all immediate post-shock quantities (with subscript 2), $S_{\rm
    sub}$, $S_{\rm tot}$, $p_{\rm inj}$ and $\eta$.
\item Update $U_{px}(p)$ and goto item (v) above.
\item In general convergence will be achieved for $K \neq 1$. However, the
correct solution is only obtained when $K=1$, which will usually require
restarting the calculation with a different initial value of
$U_{1}$ (i.e. goto item (i) above). This can be driven by a standard numerical root-finding procedure.
\end{enumerate}
The distribution function $f(x,p)$ so obtained is then a solution of both the
transport and conservation equations.
 
The flux of non-thermal particles that escape upstream of the shock
can be determined from the equation for the conservation of energy
flux
\begin{eqnarray}
\label{eq:energyFractions}
& \frac{1}{2}\rho_{2}u_{2x}^3 + \frac{\gamma_{\rm g}}{\gamma_{\rm
    g}-1}p_{\rm g2}u_{2x} + \frac{\gamma_{\rm c}}{\gamma_{\rm
    c}-1}p_{\rm c2}u_{2x} + 3p_{\rm w2}u_{2x} +
\frac{B_{2}^{2}}{4\pi}u_{2x} \nonumber \\ & = \frac{1}{2}\rho_{0}u_{0x}^3 + \frac{\gamma_{\rm g}}{\gamma_{\rm
    g}-1}p_{\rm g0}u_{0x} +  \frac{B_{0}^{2}}{4\pi}u_{0x} - F_{\rm CResc},
\end{eqnarray}
where $F_{\rm CResc}$ is the energy flux of particles escaping at the
maximum momentum from the upstream section of the fluid.

\section{Creation of secondary particles and $\gamma$-rays via proton-proton interactions}
\label{sec:appProtonProton}
In this appendix we provide equations for some of the emissivity calculations
in our models (see \citet{Orellana:2007} and \citet{Vila:2012} for further details). 

Proton-proton inelastic collisions create pions through the following
reactions:
\begin{align}
\label{eq:proton_proton_pion_production}
p+p&\rightarrow p+p+a\pi^0+b\left(\pi^++\pi^-\right)\nonumber\\[0.0cm]
p+p&\rightarrow p+n+\pi^++a\pi^0+b\left(\pi^++\pi^-\right)\\[0.0cm]
p+p&\rightarrow n+n+2\pi^++a\pi^0+b\left(\pi^++\pi^-\right)\nonumber.
\end{align}
The integers $a$ and $b$ are the pions multiplicities. They depend on
the energy of the relativistic proton approximately as
$a,b\propto E_p^{-\kappa}$ with $\kappa\sim1/4$
\citep{Mannheim:1994}. The threshold energy for the production of a
single neutral pion is
\begin{equation}
\label{eq:threshold_energy_pion_creation_pp}
E_{\rm thr} = m_{\rm p}c^2  + 2m_{\pi^0} c^2 \left(1 +
  \frac{m_{\pi^0}}{4m_{\rm p}}\right)\approx 1.22\, \mathrm{GeV},
\end{equation} 
where $m_{\rm p}$ and $m_{\pi^0}$ are the mass of the proton and the neutral pion, respectively. 

The decay of neutral pions into $\gamma$-rays is calculated as in
Appendix~A3 in \citet{Pittard:2020}. The main decay channels for the
charged pions created in proton-proton inelastic collisions are:
\begin{align}
\label{eq:charged_pion_decay}
\pi^+\rightarrow\mu^++\nu_\mu,\\[0.0cm]
\pi^-\rightarrow\mu^-+\overline{\nu}_\mu \nonumber,
\end{align}
with a branching ratio of $99.98770\pm0.00004$
\citep{Nakamura:2010}. Muons decay with a probability almost equal to
unity into a neutrino, an antineutrino, and an electron/positron:
\begin{align}
\label{eq:charged_pion_decay}
\mu^+\rightarrow e^++\nu_{\rm{e}}+\overline{\nu}_\mu\\[0.0cm]
\mu^-\rightarrow e^-+\overline{\nu}_{\rm{e}}+\nu_\mu.\nonumber
\end{align}
The charged muons decay over very short distances compared to the size
of the WCR, so the electron and positron are assumed to be injected
``on-the-spot''.

\citet{Kelner:2006} provide simple analytical formulae for the
cross-section and energy spectra of the products of inelastic
proton-proton collisions. However, their fits are dependent on the
shape of the non-thermal proton distribution. As a result we instead
use the $\delta$-functional approximation to obtain the injection
function of electrons:
\begin{equation}
\label{eq:injection_electrons_chargedpion_decay}
Q_{\rm e\pm}(E_{\rm e}) = c \frac{n_{\rm p}}{K_{\rm e}} \sigma_{\rm pp}(E_{\rm p})
N_{\rm p}(E_{\rm p}), 	
\end{equation}
where $E_{\rm e} = K_{\rm e}E_{\rm kin}$ and $K_{\rm e}$ is the
fraction of the proton energy that the electron has. For electron
production via charged pions, $K_{\rm e} = K_{\pi}/4$, where
$K_{\pi}=0.17$ is the fraction of the proton kinetic energy that the leading pion has.
$n_{\rm p}$ is the number density of thermal protons, and $N_{\rm
  p}$ is the distribution function of the non-thermal protons (units
of protons\,cm$^{-3}$\,erg$^{-1}$).

The inelastic proton-proton cross section $\sigma_{\rm{pp}}\left(E_{\rm p}\right)$
can be accurately approximated as \citep{Kelner:2006}
\begin{equation}
\label{eq:cross_section_pp}
\sigma_{\rm{pp}}\left(E_{\rm p}\right)=\left(34.3+1.88\,L+0.25\,L^2\right)\left[1-\left(\frac{E_{\rm
        th}}{E_{\rm p}}\right)^4\right]^2\,\rm{mb}.
\end{equation}

\section{Creation of secondary particles and $\gamma$-rays via
  proton-photon interactions}
\label{sec:appProtonPhoton}
Proton-photon inelastic collisions create pions and electron-positron
pairs through the following reactions:
\begin{align}
\label{eq:photopair_and_photomeson}
p+\gamma & \rightarrow p+e^-+e^+\nonumber\\
p+\gamma & \rightarrow p+a\pi^0+b\left(\pi^++\pi^-\right)\\
p+\gamma & \rightarrow n+\pi^++a\pi^0+b\left(\pi^++\pi^-\right) \nonumber.
\end{align}  
The integer coefficients $a$ and $b$ are, as before, the pion
multiplicities. In the rest frame of the proton, the photon threshold
energy for the creation of a pair is
\mbox{$\epsilon_{\rm thr}^{\prime(e)} = 2m_{\rm e}c^2\approx1$}\,MeV.
Photomeson production becomes possible when the energy of the photon
in the rest frame of the proton is larger than
\begin{equation}
\label{eq:threshold_energy_photomeson}
\epsilon_{\rm thr}^{\prime(\pi)} = m_{\pi^0} c^2\left(1 +
  \frac{m_{\pi^0}}{2m_{\rm p}}\right)\approx 145\, \mathrm{MeV},
\end{equation}  
The cross section for pair production is about two orders of magnitude
larger than that for pion production. The inelasticity
$K_{p\gamma}^{(e)}$, however, is very low, so the proton only loses a
small fraction of its energy per collision. As a result, the cooling
is completely dominated by pion production if the energy of the
photons exceeds $\epsilon_{\rm thr}^{\prime(\pi)}$.

If the cooling of pions and muons before decay is neglected, the
injection function of electron-positron pairs can be easily estimated
in the $\delta$-functional approximation as in
\citet{Atoyan:2003}. Assuming that each charged pion takes an energy
$E_\pi \approx 0.2 E_{\rm p}$, and that this energy is equally
distributed among the decay products, the energy of each
electron/positron is $E_{\rm e}\approx 0.05\,E_{\rm p}$. The injection
function of pairs is then

\begin{equation}
\label{eq:injection_pairs_photomeson_AD}	
 Q_{\rm e^{\pm}}\left(E_{\rm e^{\pm}}\right)=20 N_{\rm p}\left(20E_{\rm
    e}\right)\omega_{p\gamma}^{(\pi)}\left(20E_{\rm
    e}\right)n_{\pi^\pm}\left(20E_{\rm e}\right),        	
\end{equation} 
where $n_{\pi^\pm}$ is the mean number of charged pions created per
proton-photon collision and $\omega_{p\gamma}^{(\pi)}$ is the
collision rate. Thus the secondary electrons have the same spectral
shape as the non-thermal protons and extend up to a maximum energy of
$0.05\,E_{\rm p,max}$ where $E_{\rm p,max}$ is the maximum proton
energy.

\citet{Kelner:2008} provide simple analytical expressions for the
spectrum of gamma rays due to decay of neutral pions created in
proton-photon collisions. In terms of the distributions of
relativistic protons and target photons, the gamma-ray emissivity can
be written as
\begin{equation}
\label{eq:gamma_ray_emissivity_proton_photon_KA}
q_\gamma^{(p\gamma)}\left(E_\gamma\right)=\int_{E_p^{\rm{min}}}^{E_p^{\rm{max}}}\,\mathrm{d}E_p\int_{\epsilon_{\rm{thr}}^{\prime(\pi)}/2\gamma_p}^{\infty}\,\mathrm{d}\epsilon\,\frac{N_p(E_p)}{E_p}\,n_{\rm{ph}}(\epsilon)\,\Phi\left(\eta,\,x\right).
\end{equation} 
Here $\eta=4\epsilon E_{\rm p}/m_{\rm p}^2c^4$ and $x=E_\gamma/E_{\rm
  p}$. The function $\Phi\left(\eta,\,x\right)$ was obtained fitting
the numerical results of SOPHIA, a Monte Carlo code for the simulation
of photohadronic interactions \citep{Mucke:2000}. The function
$\Phi\left(\eta,\,x\right)$ can be approximated with an accuracy
better than 10\% by a simple analytical formula. If we define $x_\pm$ as
\begin{equation}
\label{eq:aux_function_pgamma_KA_1} 
x_\pm=\frac{1}{2(1+\eta)}\left[\eta+r^2\pm\sqrt{\left(\eta-r^2-2r\right)(\eta-r^2+2r)}\right],
\end{equation} 
then, in the range $x_-<x<x_+$,
\begin{multline}
\label{eq:aux_function_pgamma_KA_2}
\Phi_\gamma\left(\eta,\,x\right) =
  B_\gamma\,\exp\left\{-s_\gamma\left[\ln\left(\frac{x}{x_-}\right)\right]^{\delta_\gamma}\right\}
  \\
 \times \left[\ln\left(\frac{2}{1+y^2}\right)\right]^{2.5+0.4\ln(\eta/\eta_0)},
\end{multline} 
where 
\begin{equation}
\label{eq:aux_function_pgamma_KA_3}
y=\frac{x-x_-}{x_+-x_-}, 
\end{equation} 
and 
\begin{equation}
\label{eq:eta_0_pgamma}
\eta_0 = 2\frac{m_\pi}{m_p} + \frac{m_\pi^2}{m_p^2}\approx 0.313.
\end{equation} 
For $x<x_-$, the spectrum is independent of $x$,
\begin{equation}
\label{eq:aux_function_pgamma_KA_4}
\Phi_\gamma\left(\eta,\,x\right)=B_\gamma\,\left[\ln2\right]^{2.5+0.4\ln(\eta/\eta_0)},
\end{equation} 
and $\Phi_\gamma\left(\eta,\,x\right)=0$ for $x>x_+$. The parameters $B_\gamma$, $s_\gamma$ and $\delta_\gamma$ are functions of $\eta$. For values of $1.1\,\eta_0<\eta<100\,\eta_0$, these functions are tabulated in \citet{Kelner:2008}.

For power-law distributions of protons, \citet{Kelner:2008} claim that
it is more convenient to integrate over $d\eta$. i.e.
\begin{equation}
\frac{dN}{dE} = \int_{\eta_{0}}^{\infty} H(\eta,E)\,d\eta,
\end{equation}
where
\begin{equation}
H(\eta,E) = \frac{m_{\rm p}^{2}c^{4}}{4} \int_{E}^{\infty}
\frac{N_{\rm p}(E_{\rm p})}{E_{\rm p}^{2}} n_{\rm ph}\left(\frac{\eta
  m_{\rm p}^{2}c^{4}}{4 E_{\rm p}}\right)
\Psi\left(\eta,\frac{E}{E_{\rm p}}\right)\,dE_{\rm p}.
\end{equation}
In practice, we change the integration variable from $x$ to $u = {\rm
  log_{10}}x$.

\section{Comparison of electron and $\gamma$-ray emission by
  proton-proton and proton-photon interactions}
\label{sec:appPPvsPgam}
\begin{figure}
\includegraphics[width=8.0cm]{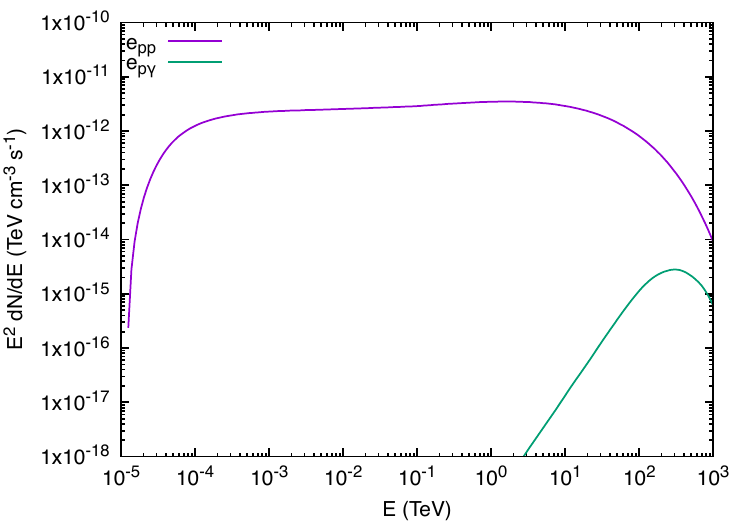}
\includegraphics[width=8.0cm]{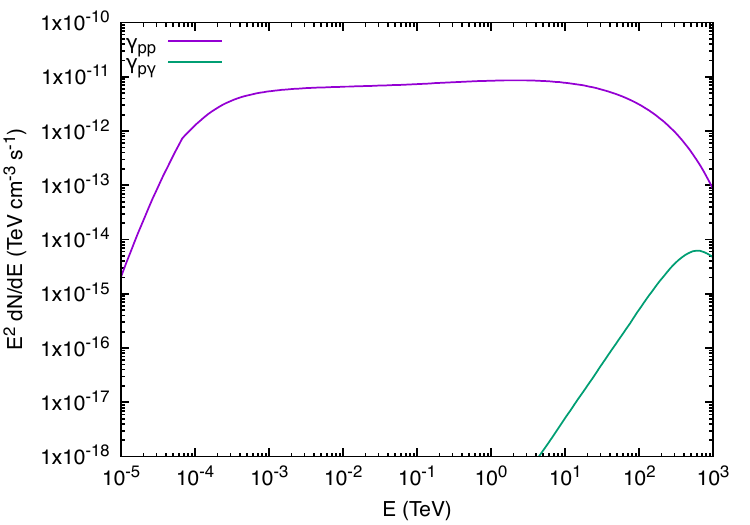}
\caption{Comparison of the electron (top) and photon (bottom)
  emissivities for proton-proton and proton-photon
  interactions. Conditions appropriate for the stagnation point of the
WCR are assumed, including a thermal proton number density of
$3.9\times10^{5}\,{\rm cm^{-3}}$ and a solid angle of
$10^{-6}$\,steradian for the disk of the O-star.}
\label{fig:pp_vs_pgam}
\end{figure}

Fig.~\ref{fig:pp_vs_pgam} shows the electron and $\gamma$-ray spectra produced
via proton-proton (pp) and proton-photon (p$\gamma$) interactions. In
each case the non-thermal protons have a distribution specified by 
\begin{equation}
J_{\rm p}(E_{\rm p}) = A E_{\rm p}^{-2} \,{\rm
  exp}\left(-\frac{E_{\rm p}}{E_{0}}\right),
\end{equation}
with $E_{\rm 0}=1000$\,TeV and $A$ set so that
\begin{equation}
\int_{\rm 1\,TeV}^{\infty} E_{\rm p}J_{\rm p} \,dE_{\rm p} = 1\,{\rm erg\,cm^{-3}}.
\end{equation} 
The thermal proton density is assumed to be $4\times10^{5}\,{\rm
  cm^{-3}}$ (as appropriate for the post-O-shock gas at the stagnation
point in the WCR). At this location the secondary star (which
dominates the photon flux) occupies a solid angle of $\approx
10^{-6}$\,steradian (the stellar radius of the O-star, $R_{\rm
  *O}=14.72\,\Rsol$). It is clear that the proton-proton emissivity is
much stronger than the proton-photon emissivity, and that
proton-photon emission can be safely ignored in this work.

\section{Synchrotron emission}
\label{sec:appSynchrotronEmission}
The synchrotron power per unit energy radiated by a single electron of
energy $E_{\rm e}$ in a vacuum is given by
\citep[e.g.][]{Ginzburg:1965,Vila:2012}
\begin{equation}
\label{eq:Psync}
P_{\rm sync}(E_{\gamma},E_{\rm e},\alpha) =
\frac{\sqrt{3}e^{3}B}{m_{\rm e}c^{2}h}\frac{E_{\gamma}}{E_{\rm
    c}}\int^{\infty}_{E_{\gamma}/E_{\rm c}} K_{5/3}(\zeta){\rm d}\zeta,
\end{equation}
where $E_{\gamma}$ is the energy of the emitted photon and
$K_{5/3}(\zeta)$ is a modified Bessel function of the second
kind. $P_{\rm sync}$ peaks sharply near the characteristic energy
\begin{equation}
E_{\rm c} = \frac{3 h e B \sin\alpha}{4\pi m_{\rm e}c}\gamma_{\rm e}^{2},
\end{equation}
where $\gamma_{\rm e}$ is the electron Lorentz factor, and the pitch
angle $\alpha$ is the angle between the magnetic field (with flux
density $B$) and the particle's momentum. For a turbulent/isotropic
magnetic field, $\sin\alpha=\sqrt{2/3}$. The integral in
Eq.~\ref{eq:Psync} can be approximated as \citep{Melrose:1980}
\begin{equation}
x \int^{\infty}_{x} K_{5/3}(\zeta){\rm d}\zeta \approx
1.85x^{1/3}e^{-x},
\end{equation}
where $x=E_{\gamma}/E_{\rm c}$ is the dimensionless energy.
 
The situation changes somewhat if instead the electron is in the presence of a cold background
plasma. In such a case the refractive index of the medium, which is
smaller than unity, reduces the beaming effect, and can greatly reduce
the synchrotron emission. This effect is know as the Tsytovitch-Razin
effect \citep[or, more generally, as the Razin
effect;][]{Tsytovitch:1951,Razin:1960}. Eq.~\ref{eq:Psync} now becomes
\citep*{Ginzburg:1965,VanLoo:2004}
\begin{equation}
\label{eq:Psync_Razin}
P_{\rm sync}(E_{\gamma},E_{\rm e},\alpha) =
\frac{\sqrt{3}e^{3}B}{m_{\rm e}c^{2}h}f\frac{E_{\gamma}}{E_{\rm
    c}^{1}}\int^{\infty}_{E_{\gamma}/E_{\rm c}^{1}} K_{5/3}(\zeta){\rm d}\zeta,
\end{equation}
where 
\begin{equation}
f = \left(1 + \frac{\nu_{0}^{2}}{\nu^{2}}\gamma_{\rm e}^{2}\right)^{-1/2},
\end{equation}
$\nu_{0} = \sqrt{n_{\rm e}e^{2}/\pi m_{\rm e}}$ is the plasma frequency, and
$n_{\rm e}$ is the number density of thermal electrons. $E_{\rm c}^{1}
= f^{3} E_{\rm c}$.
 
Eq.~\ref{eq:Psync_Razin} may be used for even mildly relativistic
electrons, provided that $\chi=\nu/\nu_{\rm b}\gtsimm 100$, where
$\nu_{\rm b} = \omega_{\rm b}/2\pi$ and
$\omega_{\rm b}=eB/m_{\rm e}c$ is the cyclotron frequency of the
electron. To capture the transition from cyclotron to synchrotron
emission, a data table of emissivity values for
$1 < \gamma < 5$ and $10^{-2} < \chi < 2\times10^{3}$ is created using the
approach given in \citet{Mahadevan:1996}.

The photon emission (photon\,s$^{-1}$\,erg$^{-1}$) is then obtained
from integrating over the volume and the distribution
of non-thermal electrons:
\begin{equation}
q_{\rm sync}(E_{\gamma}) = \frac{1}{E_{\gamma}}\int_{V} dV \sin\alpha
\int_{E_{\rm e}^{\rm
    min}}^{E_{\rm e}^{\rm max}} N_{\rm e}(E_{\rm e}) P_{\rm
  sync}(E_{\gamma},E_{\rm e},\alpha) dE_{\rm e},
\end{equation}
where $N_{\rm e}$ is the distribution function of the non-thermal electrons (units
of electrons\,cm$^{-3}$\,erg$^{-1}$).

\section{Photon-photon absorption by pair creation}
\label{sec:appPhotonPhotonAbsorption}
In CWB systems the stars provide large numbers of target photons for
electron-positron pair production with high-energy ($\sim$\,TeV)
$\gamma$-rays. The probability of absorption depends on the cosine of
the angle between the directions of the two photons, $\mu$. For a
$\gamma$-ray of energy $E$ interacting with a stellar photon of energy
$\epsilon$, the optical depth is given by
\begin{equation}
\frac{{\rm d}\tau}{{\rm d}\epsilon {\rm d}\Omega {\rm d}l} =
(1-\mu)n_{\rm ph}\sigma_{\gamma\gamma},
\end{equation}
where ${\rm d}l$ is the distance along the path of the $\gamma$-ray,
${\rm d}\Omega$ is the solid angle of the stellar surface,
and $n_{\rm ph}$ is the radiation density which is assumed to be that of a
blackbody of temperature $T_{*}$:
\begin{equation}
n_{\rm ph} = \frac{2\epsilon^{2}}{h^{3}c^{3}}\frac{1}{{\rm
    exp}(\epsilon/(kT_{*}))-1} \hspace{1cm}{\rm (ph\,cm^{-3}\,erg^{-1}\,sr^{-1})}.
\end{equation}
The cross-section depends only on $\beta=(1 - 1/s)^{1/2}$, where 
$s =\epsilon/\epsilon_{\rm min}$ and the threshold energy
\begin{equation}
\epsilon_{\rm min}=\frac{2 m_{\rm e}^{2}c^{4}}{E(1-\mu)}.
\end{equation}
The cross-section is \citep{Gould:1967}
\begin{equation}
\sigma_{\gamma\gamma}(\beta) = \frac{3}{16}\sigma_{\rm T}(1 -
\beta^{2})\left[(3-\beta^{4}){\rm ln}\left(\frac{1 + \beta}{1 -
      \beta}\right) - 2\beta(2-\beta^{2})\right]. 
\end{equation}
We follow the prescription given in \citet{Dubus:2006} to calculate
the opacity \citep*[see also][]{Romero:2010}. The integral over solid angle can be split into one over
$\mu$ and $\phi$. As noted by \citet{Dubus:2006}, the energy integral can be replaced
with a definite integral over $\beta$ between the limits [0,1] while
the integral along $l$ can be replaced with a definite integral over
the angle $\psi$ between the limits [$\psi_{0},\pi$]. The final equation for
calculating the optical depth then becomes
\begin{multline}
\tau = \int_{\psi_{0}}^{\pi} \frac{d_{0}\sin\psi_{0}}{\sin^{2}\psi}{\rm d}\psi \int_{\mu_{\rm min}}^{1} {\rm d}\mu
\int_{0}^{2\pi} {\rm d}\phi \\ \times \int_{0}^{1} \frac{2 \epsilon^{2}\sqrt{1 -
    \beta}}{\epsilon_{\rm min}}  (1-\mu)n_{\rm ph}\sigma_{\gamma\gamma}{\rm d}\beta,
\end{multline}
where $\mu_{\rm min} = (1 - R_{*}^{2}/d^{2})^{1/2}$ and $d$ is the
distance of the $\gamma$-ray to the star.

\section{Free-free absorption}
\label{sec:appFreeFreeAbsorption}
The emission from the WCR can also be absorbed by the stellar
winds. In our axisymmetric model, the line of sight to the observer
from a given patch on the WCR may pass solely through the primary or
secondary wind, or it may pass first through the secondary wind
and then move into the primary wind\footnote{If orbital motion is
  included, the WCR obtains a spiral shape, and multiple transitions
  between primary and secondary wind material may occur along each
  line of sight - see \citet{Parkin:2008}.}. 

The dot product of the line of sight vector with the
normal to the shock determines if the line of sight moves into the
primary wind or into the secondary wind. If the latter occurs the line
of sight may remain in the secondary wind, or it may intersect another
part of the WCR and then move into the primary wind. To determine if
this latter case occurs, a triangle or quadrangle facet is constructed
from each patch on the WCR. Triangular facets are constructed only for
those patches that touch the apex of the WCR, while quadrangles are
constructed for all other patches. The list of triangle and quadrangle
facets is then run through to find intersections with the line of
sight, using standard techniques \citep{Schlick:1993,Moller:1997}.

For an ionized stellar wind consisting solely of protons and electrons
the optical depth due to free-free absorption along a line of sight,
$s$, is given by
\citep[see][]{Wright:1975,Panagia:1975}\footnote{There is an
  equivalence between the geometry used by \citet{Wright:1975} and
  \citet{Panagia:1975} for the free-free absorption, and
  \citet{Dubus:2006} for the photon-photon absorption.}
\begin{equation}
\label{eq:tauFF}
{\rm d}\tau = \int_{s_{0}}^{s_{1}} n^{2} K {\rm d}s = n_{0}^{2} K
\int_{s_{0}}^{s_{1}} \frac{{\rm d}s}{s^{2} + q^{2}} =
\frac{n_{0}^{2}K}{2 q^{3}}\left[\frac{qs}{d^{2}} + \tan^{-1}\left(\frac{s}{q}\right)\right]_{s_{0}}^{s_{1}}.
\end{equation} 
Here the proton (and electron) number density is
\begin{equation}
n = \frac{n_{0}}{s^{2} + q^{2}},
\end{equation} 
with
\begin{equation}
n_{0} = \frac{\Mdot}{4\pi \vinfty m_{\rm H}}.
\end{equation} 
The line of sight ray has an impact parameter $q$ with the star, $s=0$
at the point of closest approach to the star, and the observer is at
$s=+\infty$. In Eq.~\ref{eq:tauFF}, $s_{0}$ is the starting point of
the ray and $s_{1}$ is either equal to $+\infty$ or the value obtained
at the intersection point of the WCR. In the latter case the total
opacity is obtained from $\tau = d\tau_{1} + d\tau_{2}$, where
$d\tau_{1}$ and $d\tau_{2}$ are the optical depths through each wind
along the line of sight.

In cgs units, $K$ is given by \citep{Wright:1975}
\begin{equation}
K=3.7\times10^{8}[1 - \exp(-h\nu/kT)]
g(\nu,T)T^{-1/2}\nu^{-3}\hspace{5mm}{\rm cm^{5}},
\end{equation} 
where $g$ is the Gaunt factor. $K$ is appropriately scaled for a wind
with other atomic species. In evaluating $K$ we assume that the wind
temperature is maintained at $10^{4}$\,K. Hydrogen and helium are
assumed to be singly ionized while C, N and O are assumed to be doubly
ionized. The relative number densities of H, He and CNO are assumed to
be given by $X$, $Y/4$ and $Z/14.24$.  Finally, if the wind is clumpy,
Eq.~\ref{eq:tauFF} is scaled by $1.0/f$, where $f$ is the volume
filling factor of the clumps ($f < 1.0$).

Additional absorption due to the material in the WCR can also be
included, if desired \citep[see, e.g.,][]{Parkin:2008}. However, this
addition is only likely to be significant if the WCR strongly cools,
and so is not included in the present work.

\bsp	
\label{lastpage}
\end{document}